\documentclass{jfm}
\usepackage{graphicx}
\usepackage{epstopdf, epsfig}
\usepackage{caption}
\usepackage{amsmath}
\usepackage{subcaption}
\usepackage{soul}
\usepackage{tikz,siunitx}

\usepackage{cleveref}
\usepackage{lineno}

\shorttitle{The effect of sand grain roughness on Taylor--Couette turbulence}
\shortauthor{P. Berghout, X. Zhu, R. Verzicco, D. Chung, R.J.A.M. Stevens, D. Lohse}

\title{Direct numerical simulations of Taylor--Couette turbulence: the effect of sand grain roughness}

\author{Pieter Berghout\aff{1}
  \corresp{\email{p.berghout@utwente.nl}},
  Xiaojue Zhu\aff{1},
  Daniel Chung\aff{2},  
  Roberto Verzicco\aff{1,3},  
  Richard J. A. M. Stevens\aff{1},
  Detlef Lohse\aff{1,4}}

\affiliation{\aff{1}Physics of Fluids Group and Max Planck Center Twente, MESA+ Institute and J. M. Burgers Centre for Fluid Dynamics,
University of Twente, P.O. Box 217, 7500AE Enschede, Netherlands
\aff{2}Department of Mechanical Engineering, University of Melbourne, Victoria 3010, Australia
\aff{3}Dipartimento di Ingegneria Industriale, University of Rome `Tor Vergata', Via del Politecnico 1, Roma 00133, Italy
\aff{4}Max Planck Institute for Dynamics and Self-Organisation, Am Fassberg 17, 37077 G\"{o}ttingen, Germany}

\begin{document}

\maketitle

\begin{abstract}
Progress in roughness research, mapping any given roughness geometry to its fluid dynamic behaviour, has been hampered by the lack of accurate and direct measurements of skin-friction drag, especially in open systems. The Taylor--Couette (TC) system has the benefit of being a closed system, but its potential for characterizing irregular, realistic, 3-D roughness has not been previously considered in depth. 

Here, we present direct numerical simulations (DNSs) of TC turbulence with sand grain roughness mounted on the inner cylinder. The model proposed by Scotti (\textit{Phys. Fluids}, vol. 18, 031701, 2006) has been improved to simulate a random rough surface of monodisperse sand grains, which is characterized by the equivalent sand grain height $k_s$. Taylor numbers range from $Ta = 1.0\times 10^7$(corresponding to $Re_\tau = 82$) to $Ta = 1.0\times 10^9$($Re_\tau = 635$). We focus on the influence of the roughness height $k_s^+$ in the transitionally rough regime, through simulations of TC with rough surfaces, ranging from $k_s^+=5$ up to $k_s^+ = 92$, where the superscript `$+$' indicates non-dimensionalization in viscous units.

We analyse the global response of the system, expressed both by the dimensionless angular velocity transport $Nu_\omega$ and by the friction factor $C_f$. An increase in friction with increasing roughness height is accompanied with enhanced plume ejection from the inner cylinder. Subsequently, we investigate the local response of the fluid flow over the rough surface. The equivalent sand grain roughness $k_s^+$ is calculated to be $1.33k$, where $k$ is the size of the sand grains. 
We find that the downwards shift of the logarithmic layer, due to transitionally rough sand grains exhibits remarkably similar behavior to that of the Nikuradse (\textit{VDI-Forschungsheft} 361, 1933) data of sand grain roughness in pipe flow, regardless of the Taylor number dependent constants of the logarithmic layer. 
\end{abstract}

\section{Introduction}
Many turbulent flows in nature and industry are bounded by rough boundaries. Examples are the surface of the planet with respect to geophysical flows or fouling on ship hulls with respect to open waters. Such rough boundaries strongly influence the total drag, with often adverse consequences in the form of higher transport costs. Therefore, it becomes of paramount importance to understand the physics behind such changes in drag, ultimately leading to better informed management of the problem. One key recurring question concerns the influence of the roughness topology on the drag coefficient. 

Seminal work by \cite{Nikuradse33} investigated the influence of the height of closely packed, monodisperse, sand grains in pipe flow. This work has become one of the pillars in the field. Later, a vast amount of research was carried out to study the influence of roughness on the canonical systems of turbulence -- pipe, channel, and boundary layer flows -- aiming for a better understanding of the roughness effects on turbulent flows \cite{ligrani86,schultz07,chan15,busse17}; also see \cite{jimenez04}, and \cite{flack10} for comprehensive reviews.

Next to pipe flow, Taylor--Couette (TC) flow -- the flow between two coaxial, independently rotating cylinders -- is another canonical system in turbulence \citep{grossmann16}. Closely related to its `twin' of Rayleigh--B\'enard (RB) turbulence \citep{busse12, eckhardt07}, it serves as an ideal system to study the interaction between boundary layer and bulk flow. Very long spatial transients, as found in open systems, are bypassed by the circumferential restrictions. Since the domain is closed, global balances can be easily derived and monitored, giving room for extensive comparison between theory, experiments and simulations. Further, the streamwise curvature effects find many applications in industry. For these reasons, we set out to investigate the effects of roughness on the turbulent fluid flow in the TC system.

Over the last century, much work has been carried out with the aim of understanding the effect of the roughness topology on fluid flow. One of the consequences of roughness is the change of the wall drag, which can be expressed as a shift of the mean streamwise velocity profile $\Delta U^+ \equiv (U_{s}-U_{r})/u_{\tau}$, where $\Delta U^+$ is known as the Hama roughness function \citep{hama54} and $U_s, U_r$ are the smooth-wall and the rough-wall mean streamwise velocities, respectively. \cite{clauser54} and \cite{hama54} already observed that roughness effects are confined to the inner region of the boundary layer. Namely that the velocity deficit occurs in the log layer. \cite{schockling06} and \cite{schultz07} later found strong support for this notion. The shift in the mean streamwise velocity profile $U(y)$ then becomes \citep{pope2000}
\begin{equation}
u^+(y^+) = \frac{1}{\kappa}\log{y^+} + A - \Delta u^+
\end{equation} 
As usual, the superscript `+' indicates a scaling in viscous units (e.g. length $y^+=yu_{\tau}/\nu$ and velocity $u^+=u/u_{\tau}$) and $u_{\tau}$ is the friction velocity, $u_{\tau} = \sqrt[]{\tau_w / \rho}$ with $\tau_w$ being the total tangential stress at the wall, and $\rho$ the fluid density. Note that in this representation, the skin-friction coefficient $C_f$ is related to the friction velocity by $C_f = 2(u_{\tau}/U_0)^2$, where $U_0$ is the centerline velocity \citep{pope2000}. It has been found that for TC turbulence $\kappa$ and $A$ are not constant anymore, but are functions of the inner cylinder Reynolds number $Re_i$ at least until $Re_i=10^6$ \citep{huisman13}. Therefore, for TC we here suggest the generalization;
\begin{equation}
u^+(y^+) = f_1(Re_i)\log(y^+)+f_2(Re_i) - \Delta u^+
\end{equation}
with $f_1(Re_i)$ and $f_2(Re_i)$ being unknown functions.
  The questions now are: $i)$ How does $\Delta u^+$ depend on the parameters that characterize the surface geometry. $ii)$ Can $\Delta u^+$ be generalized to other flows.

   Although many parameters influence the Hama roughness function $\Delta u^+$ \citep{schlichting36,musker80,napoli08,chan15,mcdonald16}, the most relevant parameter is the characteristic height of the roughness $k^+$. In a regime in which the pressure forces dominate the drag force, any surface can be collapsed onto the Nikuradse data by rescaling the roughness height to the so-called `equivalent sandgrain roughness height' $k_s^+$.  \cite{Nikuradse33} found that three regimes of the characteristic roughness height $k_s^+$ can be differentiated with respect to the effect of roughness \citep{flack12}. For $k_s^+\lesssim 5$, the rough wall appears to be hydrodynamically smooth and the roughness function $\Delta u^+$ goes to zero. For $k_s^+ \gtrsim 70$, the wall drag scales quadratically with the fluid velocity and the friction factor $C_f$ is independent of the Reynolds number, indicating that hydrodynamic pressure drag (also called form drag) on the roughness dominates the total drag. The transitionally rough regime is in between these two regimes. Where in the fully rough regime, a surface is fully determined by $k_s^+$ to give a collapse onto the fully rough asymptote \citep{flack10}, in the transitionally rough regime, any unique surface corresponds to a unique roughness function. This can be attributed to the delicate interplay between pressure drag, viscous drag, and the weakening of the so-called viscous generation cycle \citep{jimenez04}.
   
An intriguing feature of the data from \cite{Nikuradse33} is at $k_s^+\approx 5$, where roughness effects suddenly result in an inflectional increase of $\Delta u^+$, as compared to the gradual increase of the roughness function found by \cite{colebrook39} who extracted an empirical relationship from many industrial surfaces \citep{schockling06}. Later, this inflectional behavior was also observed for tightly packed spheres \citep{ligrani86}, honed surfaces \citep{schockling06}, and grit-blasted surfaces \citep{thakkar18}. \cite{chanbraun11} had too few points to find the inflectional behavior; however, their two simulations of monodisperse spheres in regular arrangement collapsed on the Nikuradse curve. In the Moody \citep{moody44} representation, this inflectional behavior manifests itself as a dip in the friction factor $C_f$, leading to a significantly lower drag coefficient ($\approx 20\%$ \citep{bradshaw2000}) in comparison to the monotonic behavior of \cite{colebrook39}, on which the original Moody diagram is based. Although it is proposed that the inflectional behavior has to do with the monodispersity of the roughness leading to a critical Reynolds number at which the elements become active \citep{jimenez04}, recent simulations by \cite{thakkar18} for a polydisperse surface (containing irregularities with a range of sizes) also show this inflectional behavior. In a broader sense, the DNS by \cite{thakkar18} are interesting since they show, for the first time, a surface that very closely and quantitatively resembles the \cite{Nikuradse33} roughness function in all regimes, the authors found $k_{s}^+=0.87k_{rms}$.

Regarding TC flow, only a few studies have looked at the effect of roughness \citep{cadot97, berg2003}. Recently, the effect of regular roughness on TC turbulence has also been investigated by means of DNS \citep{zhu16,zhu17,zhu18}. \cite{zhu16} looked at the effect of axisymmetric grooves on the torque scaling, boundary layer thickness, and plume ejections. They find that enhanced plume ejection from the roughness tips can lead to an ultimate torque scaling exponent of $Nu \propto Ta^{0.5}$, although for higher $Ta$ the exponent saturates back to the ultimate scaling effective exponent of $0.38$. \cite{zhu17} then simulate transverse bar roughness elements on the inner cylinder to disentangle the separate effects of viscosity and pressure, and find that the ultimate torque scaling exponent of $Nu \propto Ta^{0.5}$ is only possible when the pressure forces dominate at the rough boundary \citep{zhu18}. 

In contrast to the above mentioned previous work, in which the roughness consisted of well-defined transverse bars with constant distance and heights \citep{zhu17,zhu18}, in this research we set out to investigate the effect of \textit{irregular}, \textit{monodisperse} roughness, resembling the sand grain roughness reported by \cite{Nikuradse33}. We model the roughness as randomly rotated and semi-randomly translated ellipsoids of constant volume and aspect ratio, based on the roughness model (subgrid-scale) of \cite{Scotti06}. Previously, a fully resolved version of the model by \cite{Scotti06} was also used for large-eddy simulations in channel flow \citep{yuan14}. Taylor numbers in our DNS range from $Ta=1.0\times 10^7$($Re_\tau = 82$) to $Ta=1.0 \times 10^9$($Re_\tau = 635$); therefore, we capture both classical (laminar-like boundary layers) and ultimate (turbulent boundary layers) regimes \citep{rodolfo14,grossmann16,zhu18}. Moreover, whereas previous research on roughness in TC flow focussed on the torque scaling, we now look at the effects of the roughness height on the Hama roughness function $\Delta U^+$ in the transitionally rough and fully rough regimes, ranging from $k_s^+= 5$ to $k_s^+= 92$.    

This manuscript is structured as follows. In \S \ref{TCflow} \& \S \ref{numerics}, we elaborate on the TC setup, the roughness model and the numerical procedure. In \S \ref{hama}, we study the velocity profiles and present the effects of the roughness height on the Hama roughness function. In \S \ref{global}, we present the global response of the system. In \S \ref{local} we study the flow structures. 
In \S \ref{sub} the fluctuations close to the roughness are studied and in \S \ref{profiles} we present radial profiles of various other quantities.
Finally, in section \S \ref{conclusions} we draw our conclusions and propose future research directions.

\section{Taylor--Couette flow}
\label{TCflow}
\begin{figure}
\centering
\includegraphics[width=1.0\linewidth]{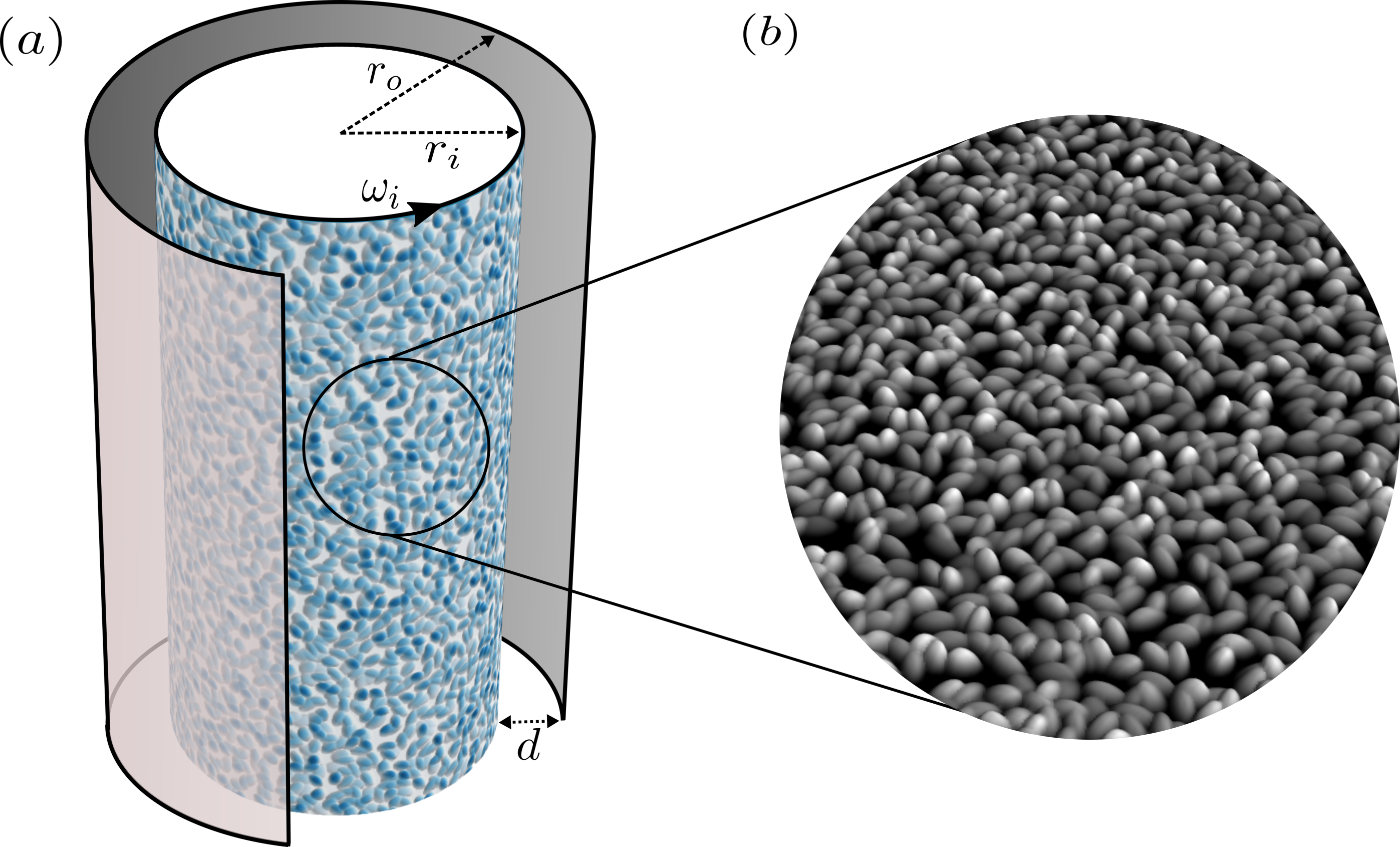}
\caption{Illustration of the Taylor--Couette setup. $(a)$ Taylor--Couette setup with inner cylinder sand grain roughness. $\omega_i$ is the inner cylinder angular rotation rate, $r_i$ is the inner cylinder radius, $r_o$ is the outer cylinder radius, and $d$ is the gap width. $(b)$ A more detailed and accurate zoom of the sand grain roughness that is modeled. The outer cylinder is stationary and smooth.}
\label{fig:TC_roughness}
\end{figure}
The TC setup, as shown in figure \ref{fig:TC_roughness}, comprises independently co- or counter-rotating concentric cylinders around their vertical axis. The flow, driven by the shear on both of the cylinders, fills the gap between the cylinders. $r_i$ is the inner cylinder radius, $r_o$ is the outer cylinder radius, and the radius ratio is defined as $\eta = r_i/r_o$. For this research, we set $\eta \approx 0.714$, in accordance with the experimental T$^3$C setup \citep{huisman13}, and previous simulations \citep{zhu17}. $\Gamma = L/d$ is the aspect ratio, where $L$ is the height of the cylinders, and $d = r_o-r_i=0.4r_i$ is the gap width. Here, $\Gamma \approx 2$ such that one pair of Taylor vortices fits in the gap, and the mean flow statistics become independent of the aspect ratio \citep{rodolfo15}. In the azimuthal direction we employ a rotational symmetry of order $6$ to save on computational expense such that the streamwise aspect ratio of our simulations becomes $L_\theta /d = (r_i 2\pi / 6)/d = 2.62$. \cite{brauckmann13} and \cite{rodolfo15} found that this reduction of the streamwise extent does not effect the mean flow statistics. This gives $L_\theta /(d/2) = 5.24$ and $L / (d/2) \approx 4.0$. 

To maintain convenient boundary conditions, we solve the Navier--Stokes (NS) equations in a reference frame rotating with the inner cylinder ($\omega_i \textbf{e}_z$) The NS equations in that reference frame become

\begin{equation}
\label{eq:NSur}
\partial_{\hat{t}} \hat{u}_{r} +  \hat{\mathbf{u}} \cdot \hat{\nabla} \hat{u}_r - \frac{\hat{u}_\theta^2}{\hat{r}} = - \partial_{\hat{r}} \hat{P} + \frac{f(\eta)}{Ta^{1/2}}(\hat\nabla^2 \hat{u}_r - \frac{\hat{u}_r}{\hat{r}^2} - \frac{2}{\hat{r}^2}\partial_{\theta} \hat{u}_\theta)-Ro^{-1}\hat{u}_\theta
\end{equation}
\begin{equation}
\label{eq:NSth}
\partial_{\hat{t}} \hat{u}_{\theta} +  \hat{\mathbf{u}} \cdot \hat{\nabla} \hat{u}_\theta - \frac{\hat{u}_\theta \hat{u}_r}{\hat{r}} = - \frac{1}{\hat{r}}\partial_{\hat{\theta}} \hat{P} + \frac{f(\eta)}{Ta^{1/2}}(\hat\nabla^2 \hat{u}_\theta - \frac{\hat{u}_\theta}{\hat{r}^2} + \frac{2}{\hat{r}^2}\partial_{\theta} \hat{u}_r)+Ro^{-1}\hat{u}_r
\end{equation}
\begin{equation}
\label{eq:NSz}
\partial_{\hat{t}} \hat{u}_{z} +  \hat{\mathbf{u}} \cdot \hat{\nabla} \hat{u}_z = - \partial_{\hat{z}} \hat{P} + \frac{f(\eta)}{Ta^{1/2}}(\hat\nabla^2 \hat{u}_z)
\end{equation}
\begin{equation}
\label{eq:in}
\hat{\nabla} \cdot \hat{\mathbf{u}} = 0
\end{equation}
with no-slip boundary conditions $u_{\theta}|_{r=r_i}=0$, $u_{\theta}|_{r=r_o}=r_o(\omega_o-\omega_i)$. Equation (\ref{eq:in}) expresses the incompressible restriction. Hatted symbols indicate the respective dimensionless variables, with $\mathbf{u} = r_i|\omega_i-\omega_o|\hat{\mathbf{u}}$,
 $r = d\hat{r}$ and $t=\frac{d}{r_i|\omega_i-\omega_o|}\hat{t}$. We define $\frac{f(\eta)}{Ta^{1/2}}=Re^{-1}$ where $f(\eta)$ is the so-called `geometric Prandtl' number \citep{eckhardt07}:
\begin{equation}
\label{eq:fn}
f(\eta)=\frac{(1+\eta)^3}{8\eta^2}
\end{equation}
here $f(0.714)\approx 1.23$ and $Ta$ is the Taylor number, which is a measure of the driving strength of the system,
\begin{equation}
\label{eq:Ta}
Ta=\frac{(1+\eta)^4}{64\eta^2}\frac{(r_o-r_i)^2(r_i+r_o)^2(\omega_i-\omega_o)^2}{\nu^2}.
\end{equation}
Note that the pressure in the equations above represents the `reduced pressure' that incorporates the centrifugal term; $\hat{P} = p' - \frac{\omega_i^2d^2\hat{r}^2}{2r^2_i|\omega_i-\omega_o|^2}\mathbf{e}_r$ with $p = \rho r_i^2|\omega_i-\omega_o|^2 p'$ 
and $p$ is the physical pressure. It is directly clear that the centrifugal force in TC flow is analogous to the gravitational force in RB flow \citep{eckhardt07}. The final term on the right-hand side of eq. (\ref{eq:NSur}) and (\ref{eq:NSth}) gives the Coriolis force, with $Ro^{-1}$ being the inverse Rossby number
\begin{equation}
\label{eq:ro}
Ro^{-1}=\frac{2\omega_id}{r_i|\omega_i-\omega_o|}.
\end{equation}     
Analogous to RB flow, the global response of TC flow can be expressed in terms of a Nusselt number. In the former, the Nusselt number expresses the dimensionless conserved heat flux, whereas in the latter the Nusselt number expresses the dimensionless conserved angular velocity flux $J^\omega$, calculated by:
\begin{equation}
\label{eq:jom}
J^\omega = r^3(\langle u_r\omega\rangle_{z,\theta,t} - \nu \partial_r \langle \omega \rangle_{z,\theta,t})
\end{equation}
with the laminar flux given by $J^\omega_{lam}=2\nu r_i^2 r_o^2\frac{\omega_1-\omega_2}{r_2^2-r_1^2}$ where $\nu$ is the kinematic viscosity and $\langle . \rangle_{z,\theta,t}$ indicates averaging over the spatial directions $z$, $\theta$ and time $t$.
For incompressible flows, it can be derived from the NS equations that $J^\omega$ is conserved in the radial direction, $\partial_r J^\omega=0$ \citep{eckhardt07}. In both cases, the values are made dimensionless by their respective laminar, conducting, values. For TC Nusselt becomes:
 
\begin{equation}
\label{eq:nu}
Nu_\omega = \frac{J^\omega}{J^\omega_{lam}}
\end{equation}
The angular velocity flux $J^\omega$ can easily be written in terms of the torque $\mathcal{T}$ on any of the cylinders: $J^\omega=\mathcal{T}(2\pi L \rho)^{-1}$ with $\rho$ being the fluid density. Consequently, the shear stress on the inner cylinder $\tau_{w,i}$ is related to the angular velocity flux by $\tau_{w,i}=\frac{\rho J^\omega}{r_i^2}$. 

Since part of our endeavour is to compare the effects of sand grain roughness on TC turbulence with the effect of sand grain roughness in other canonical systems (e.g. pipe flow), where the use of $Nu_\omega$ is not conventional, we choose to also present the global response in terms of the friction factor $C_f$. Here we follow \cite{lathrop92a}, and define $C_f\equiv G / Re_i^2$, where $G$ is the dimensionless torque $G = \mathcal{T}/(\rho \nu^2 L)$ and $Re_i$ is the inner cylinder Reynolds number $Re_i = \frac{r_i\omega_i d}{\nu}$. The translation between $Nu_\omega$ and $C_f$ is straightforward, namely
\begin{equation}
\label{eq:cf}
C_f \equiv \frac{2\pi \tau_{w,i}}{\rho d^2 |\omega_i - \omega_o|^2}= 2\pi Nu_\omega J_{lam}^\omega (\nu Re_i)^{-2}
\end{equation}
Note that one can also define $C_f \equiv 2\tau_w / \rho (r_i \omega_i)^2 = \frac{(1-\eta)^2}{\pi \eta^2}\frac{G}{Re_i^2}$, which is different from eq. (\ref{eq:cf}) by a factor which depends on the radius ratio $\eta$ \citep{lathrop92a}.    \\
 
\section{Numerical procedure}\label{numerics}
\subsection{Roughness model}
\begin{figure}
\centering
\includegraphics[width=1.0\linewidth]{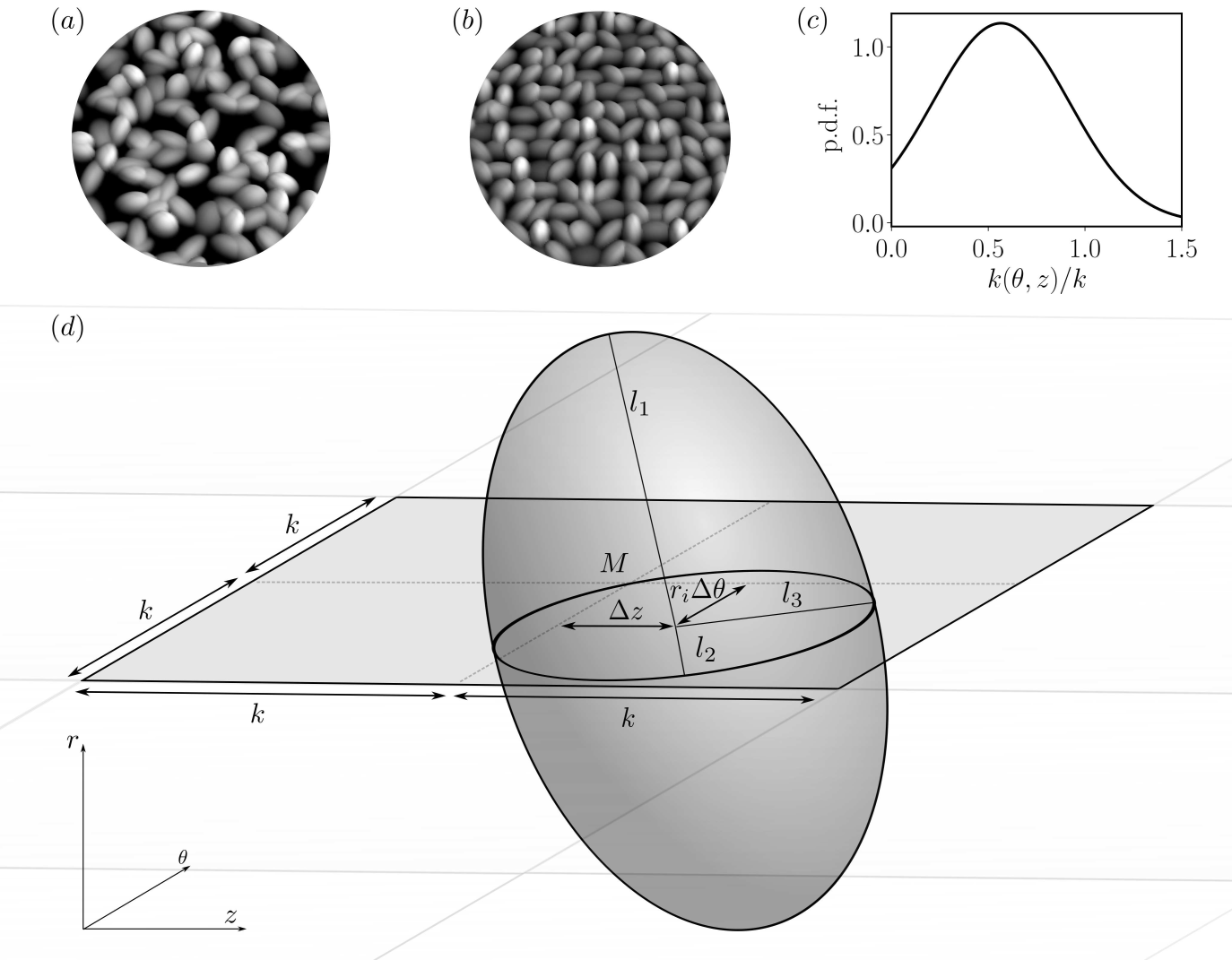}
\caption{($a$) Visual comparison between a surface with translational degrees of freedom as employed presently and ($b$) the original model by \cite{Scotti06} without translational degrees of freedom. ($c$) Probability density function (p.d.f.) of the surface height $k(\theta,z)/k$ distribution of a rough surface with $952$ roughness elements (B2). For the statistics quantifying the characteristics of all rough surfaces used in this study, we refer the reader to the Appendix, Table \ref{table:surface}. ($d$) Schematic of an ellipsoidal building block of the rough surface. Every rectangular tile of size $2k\times2k$ contains exactly one ellipsoid. $M$ indicates the center of the tile. The radii $l_1=2.0k, l_2=1.4k, l_3=1.0k$ are kept constant for each ellipsoid to maintain a \textit{monodisperse} rough surface. Randomness is ensured by giving the ellipsoid $5$ degrees of freedom; $2$ translational shifts of the center of the ellipsoid from the center of the tile $M$, ($\Delta z$ and $r_i \Delta \theta$) and $3$ rotational degrees of freedom ($\alpha_1, \alpha_2, \alpha_3$) from ($r,\theta,z$) to ($l_1,l_2,l_3$). We also employ a constant translation of the center of the ellipsoid in the radial direction, with $\Delta r  = -0.5k$ from $r=r_i$. }
\label{fig:ellipsepatch}
\end{figure}
Figure \ref{fig:ellipsepatch} exhibits the setup of the `virtual' sand grain roughness model that is used in this research. The inner cylinder is divided up into square tiles of size $2k\times2k$, each tile containing exactly $1$ ellipsoid. The height $L$ is slightly varied ($0.85d\pm0.03d$) to ensure that an integer amount of tiles fits into the domain. Unlike in the original model by \cite{Scotti06}, we also introduce a random translation of the center of the ellipsoid by applying $r_i\Delta \theta$ and $\Delta z$, where $r_i\Delta\theta$, and $\Delta z$ are random uniform translations from the center of the $2k\times2k$ tile. This random translation allows for the surface to be more irregular and as such to relate more closely to a realistic sand grain surface. As also introduced by \cite{Scotti06}, we employ a constant translation of the center of the ellipsoid in the radial direction, with $\Delta r  = -0.5k$ from $r=r_i$. It can be seen from figure \ref{fig:TC_roughness}($a$) that part of the cylinder ($\approx 15\%$) is not covered by rough elements. 

\subsection{Numerical method}
The NS equations are spatially discretized by using a central second-order finite-difference scheme and solved in cylindrical coordinates by means of a semi-implicit procedure \citep{verzicco96, vdpoel15}. The staggered grid is homogeneous in both the spanwise and streamwise directions (the axial and azimuthal directions respectively). 

The wall-normal grid consists of a double cosine (Chebychev-type) grid stretching. Below the maximum roughness height, we employ a cosine stretching such that the maximum grid spacing is always smaller than $0.5$ times the viscous length scale. In the bulk of the fluid, we employ a second stretching, such that the maximum grid spacing in the bulk is approximately $1.5$ times the viscous length scale. The minimum grid spacing is thus located at the position of the maximum roughness height, where we expect the highest shear stress, see table \ref{table:sim_param} for the exact values. 

Time advancement is performed by using a fractional-step third-order Runge--Kutta scheme in combination with a Crank--Nicolson scheme for the implicit terms. The Courant--Friedrichs--Lewy (CFL) ($\frac{u\Delta t}{\Delta x}<0.8$) time-step constraint for the non-linear terms is enforced to ensure stability.

Sand grain roughness is implemented in the code by an immersed boundary method (IBM) \citep{fadlun00}. In the IBM, the boundary conditions are enforced by adding a body force $\mathbf{f}$ to the momentum equations (\ref{eq:NSur}) - (\ref{eq:NSz}). A regular, non-body fitting, mesh can thus be used, even though the rough boundary has a very complex geometry. We perform interpolation in the spatial direction preferentially to the normal surface vector to transfer the boundary conditions to the momentum equations. The IB has been validated previously \citep{fadlun00, iaccarino03, stringano06, zhu16, zhu17, zhu18}. 

Simulations are run until they become statistically stationary, such that Nusselt $Nu_\omega $ number remains constant to within $\approx 1\%$, which requires $\hat{t}\approx 200$ . Thereafter, we gather statistics until the results converge, which requires $\hat{t}\approx 50$. The resolution constraints of the domain are typically derived from the literature and are based on the grid spacing in `+' (viscous) units. Grid independence checks of the time averaged statistics are performed by ensuring that $Nu_\omega$ remains constant with increasing grid resolution in all directions and presented along with the results in Table \ref{table:sim_param}. 

\clearpage
\section{Results}\label{results}
\begin{table}
\centering
\begin{tabular}{lll|lll|lllllllllll}
Case & $Ta$ & $k/d$ & $n_{\theta}\times n_z$ & $N_{ell}$ & $N_{\theta} \times N_z \times N_{r}$& $C_f$ & $Nu_\omega$ &  $k_{s}^+$ & $Re_{\tau}$   & $r_i^{+}\Delta \theta$  & $\Delta r^{+}$ \\
\multicolumn{8}{c}{} \\
AS & $1.0\times10^7$ & -- 	   & --			  & --  & $280\times 240\times 256$ & $0.161$ & $6.41$ &  $0.00$  & $81.9$ & $1.51$ & $0.26$ \\
A1 & $1.0\times10^7$ & $0.022$ & $59\times48$ & $64$  & $472\times 384\times 256$ & $0.170$ & $6.79$ &  $4.96$   & $84.3$ & $1.10$ & $0.22$ \\
A2 & $1.0\times10^7$ & $0.038$ & $34\times28$ & $64$  & $272\times 224\times 300$ & $0.182$ & $7.31$ &  $8.93$   & $87.4$ & $1.60$ & $0.20$ \\
A3 & $1.0\times10^7$ & $0.055$ & $24\times20$ & $144$ & $288\times 240\times 300$ & $0.191$ & $7.63$ &  $12.93$  & $89.3$ & $1.55$ & $0.22$ \\
A4 & $1.0\times10^7$ & $0.073$ & $18\times15$ & $256$ & $288\times 240\times 400$ & $0.208$ & $8.30$ &  $18.00$  & $93.2$ & $1.62$ & $0.20$ \\
\multicolumn{8}{c}{} \\
BS & $5.0\times10^7$ & --      & --		  & --  & $280\times 240\times 448$ & $0.098$ &$8.78$     &  $0.00$  & $143.3$ & $2.67$ & $0.18$\\
B1 & $5.0\times10^7$ & $0.022$ & $60\times48$& $64$  & $720\times 576\times 600$ & $0.107$ &$9.58$  &  $8.86$  & $149.7$ & $1.11$ & $0.17$\\
B2 & $5.0\times10^7$ & $0.038$ & $34\times28$& $64$  & $272\times 224\times 600$ & $0.124$ &$11.06$ &  $16.43$ & $160.9$ & $3.08$ & $0.19$\\
B3 & $5.0\times10^7$ & $0.055$ & $24\times20$& $144$ & $288\times 240\times 600$ & $0.136$ &$12.12$ &  $24.49$ & $168.4$ & $3.06$ & $0.20$\\
BY& $5.0\times10^7$  & $0.055$ & $24\times20$& $144$ & $288\times 240\times 600$ & $0.136$ &$12.20$  &  $24.57$ & $169.0$ & $3.06$ & $0.20$\\
B4 & $5.0\times10^7$ & $0.073$ & $18\times15$& $256$ & $288\times 240\times 600$ & $0.148$ &$13.24$ &  $33.99$ & $176.0$ & $3.20$ & $0.22$\\
B5 & $5.0\times10^7$ & $0.087$ & $14\times12$& $400$ & $280\times 240\times 600$ & $0.155$ &$13.84$ &  $41.71$ & $180.0$ & $3.37$ & $0.23$\\
\multicolumn{8}{c}{} \\
CS & $5.0\times10^8$ & -- & --	           & --  & $512\times 512\times 640$ &$0.060$ &$16.94$ &  $0.00$  & $354.0$ & $3.62$ & $0.23$\\
C1 & $5.0\times10^8$ & $0.019$ & $68\times56$  & $144$ & $816\times 672\times 800$ &$0.076$ &$21.48$ &  $20.39$  & $398.6$ & $2.56$ & $0.29$\\
C2 & $5.0\times10^8$ & $0.026$ & $50\times40$  & $256$ & $800\times 640\times 800$ &$0.084$ &$23.66$ &  $29.11$  & $418.4$ & $2.74$ & $0.26$\\
C3 & $5.0\times10^8$ & $0.034$ & $38\times32$  & $256$ & $608\times 512\times 800$ &$0.091$ &$25.77$ &  $39.95$  & $436.6$ & $3.76$ & $0.23$\\
C4 & $5.0\times10^8$ & $0.041$ & $32\times26$  & $256$ & $512\times 416\times 800$ &$0.096$ &$26.98$ &  $48.55$  & $446.7$ & $4.57$ & $0.28$\\
CX & $5.0\times10^8$ & $0.041$ & $32\times26$  & $256$ & $512\times 416\times 1000$&$0.096$ &$27.01$ &  $48.66$  & $447.0$ & $4.57$ & $0.20$\\
C5 & $5.0\times10^8$ & $0.047$ & $28\times23$  & $324$ & $504\times 414\times 800$ &$0.101$ &$28.50$ &  $56.98$  & $459.2$ & $4.77$ & $0.35$\\
\multicolumn{8}{c}{} \\
DS & $1.0\times10^9$ & -- & --		  & --  & $512\times 512\times 640$ & $0.054$ & $21.70$    &  $0.00$  & $476.5$  & $4.87$ & $0.30$\\
D1 & $1.0\times10^9$ & $0.026$ & $50\times40$& $256$ & $800\times 640\times 1000$& $0.080$ & $31.81$ &  $40.13$ & $576.8$  & $3.78$ & $0.27$\\
D2 & $1.0\times10^9$ & $0.034$ & $38\times32$& $400$ & $760\times 640\times 1000$& $0.086$ & $34.20$ &  $54.74$ & $598.2$  & $4.12$ & $0.29$\\
D3 & $1.0\times10^9$ & $0.041$ & $32\times26$& $484$ & $704\times 572\times 1200$& $0.089$ & $35.75$ &  $66.47$ & $611.6$  & $4.55$ & $0.23$\\
D4 & $1.0\times10^9$ & $0.047$ & $28\times23$& $784$ & $784\times 644\times 1000$& $0.094$ & $37.45$ &  $77.66$ & $625.9$  & $4.18$ & $0.45$\\
D5 & $1.0\times10^9$ & $0.055$ & $24\times20$& $1024$ & $768\times 640\times 1000$& $0.097$ & $38.54$&  $91.95$ & $635.0$  & $4.33$ & $0.46$\\
\end{tabular}
\caption{Input parameters, numerical resolution, and global response of the simulations. We run $4$ sets of simulations, which are separated by an empty horizontal line. Within every set we keep $Ta$ constant. $n_{\theta}\times n_{z}$ gives the number of ellipsoids in the streamwise ($\theta$) and spanwise ($z$) directions, respectively. $N_{ell}$ expresses the resolution ($N_{\theta}\times N_z$) per elementary building block of size $2k\times 2k$. $N_{\theta} \times N_z \times N_{r}$ presents the total resolution of the computational domain. $C_f$ is the friction factor. $Nu_\omega$ is the dimensionless torque. $k_{s}^+=1.33k^+$ is the equivalent sandgrain roughness height in viscous units, where $k^+$ is the size of the sandgrains (ellipsoids with axes $2k\times 1.4k \times k$, see figure \ref{fig:ellipsepatch}), also in viscous units. $Re_{\tau}$ is the friction Reynolds number, $Re_\tau = (d/2) u_\tau/\nu$. $r_i^{+} \Delta \theta$ is the grid spacing in the streamwise and spanwise directions in viscous units, and $\Delta r^{+}$ is the minimal grid spacing, at the maximum roughness height, in the wall-normal direction in viscous units.}
\label{table:sim_param}
\end{table}

\subsection{Roughness function}\label{hama}
\begin{figure}
\centering
\begin{subfigure}{.50\textwidth}
  \centering
  \includegraphics[width=0.94\linewidth]{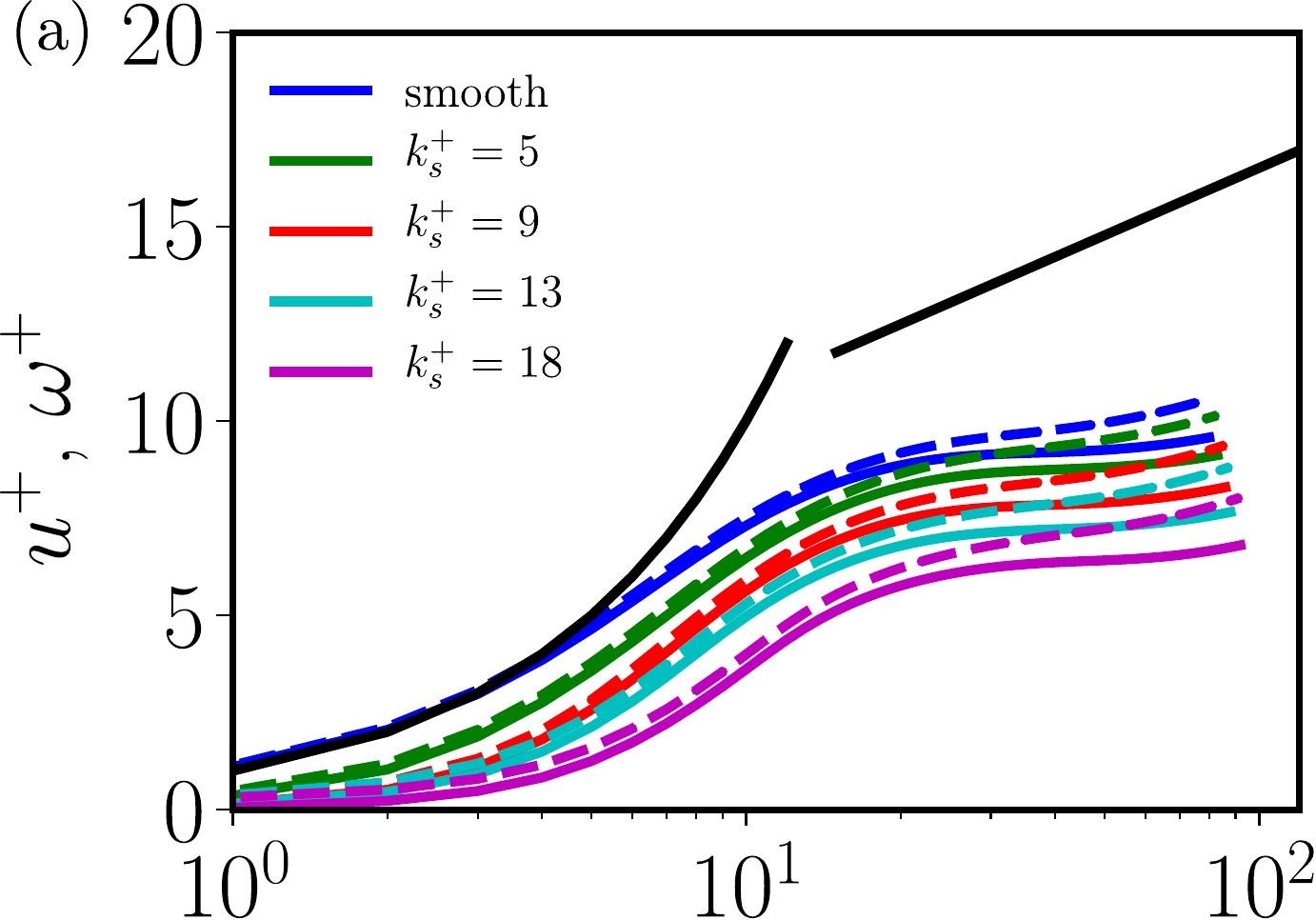}
\end{subfigure}%
\begin{subfigure}{.50\textwidth}
  \centering
  \includegraphics[width=0.94\linewidth]{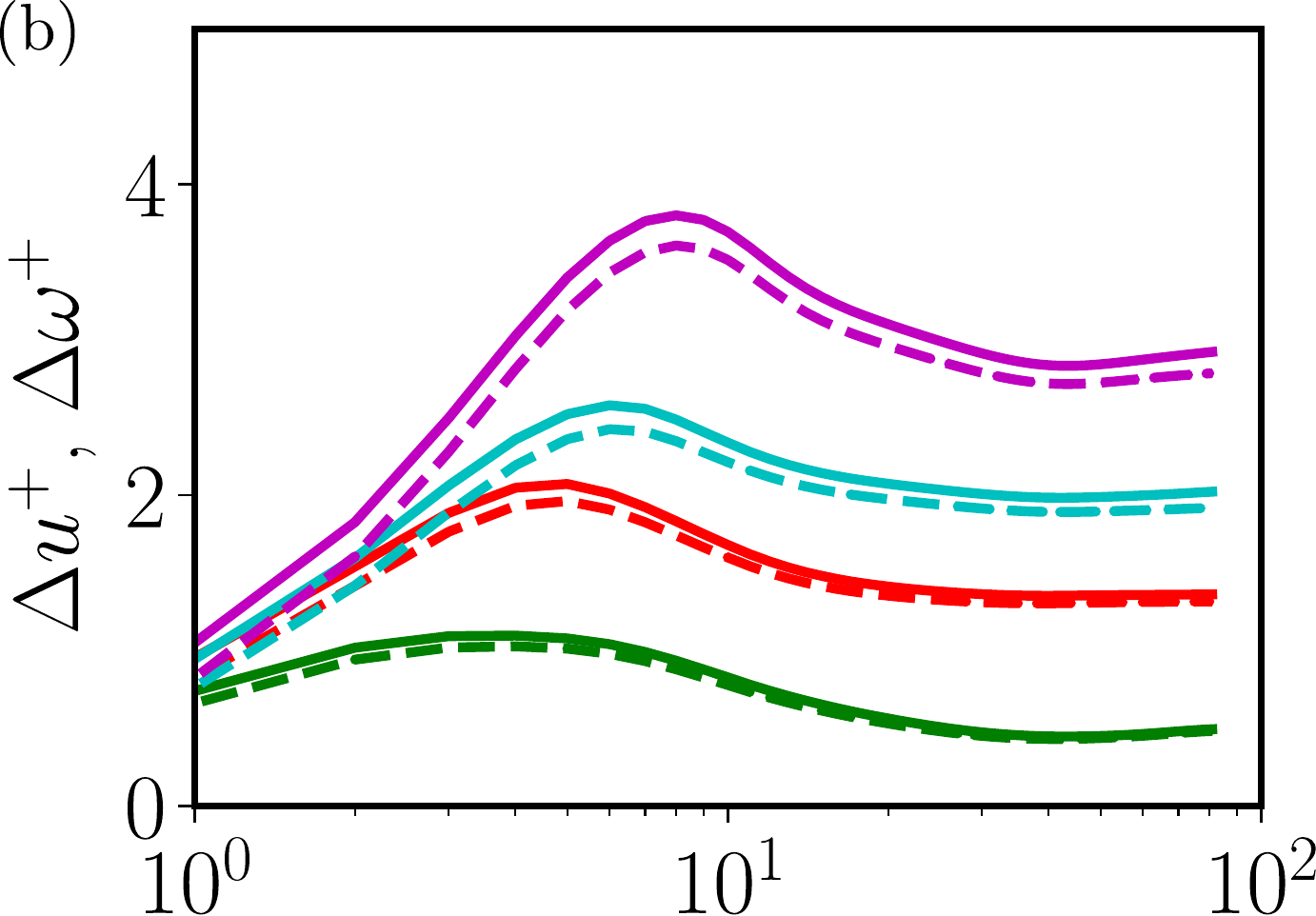}
\end{subfigure}
\begin{subfigure}{.50\textwidth}
  \centering
  \includegraphics[width=0.94\linewidth]{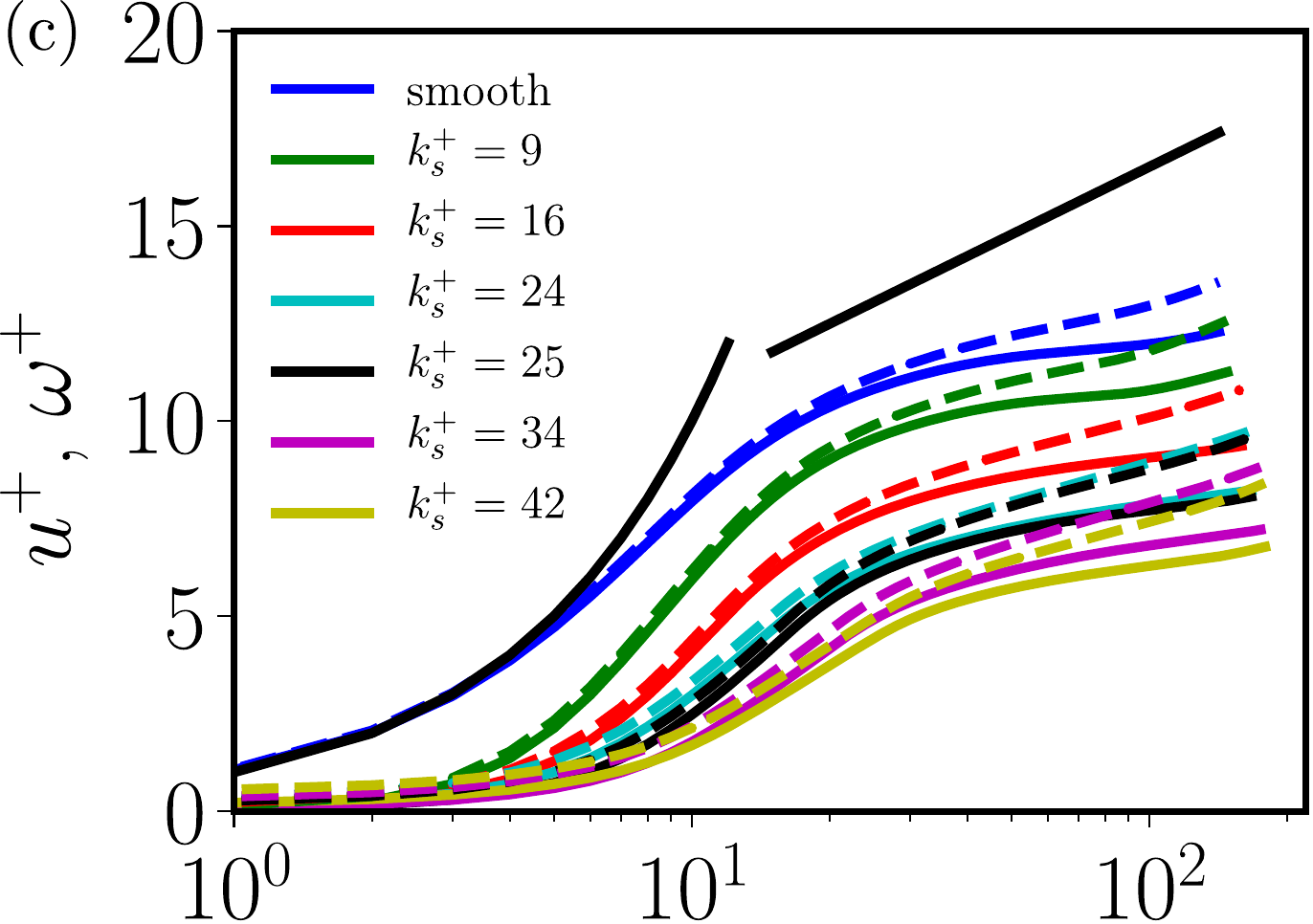}
\end{subfigure}%
\begin{subfigure}{.50\textwidth}
  \centering
  \includegraphics[width=0.93\linewidth]{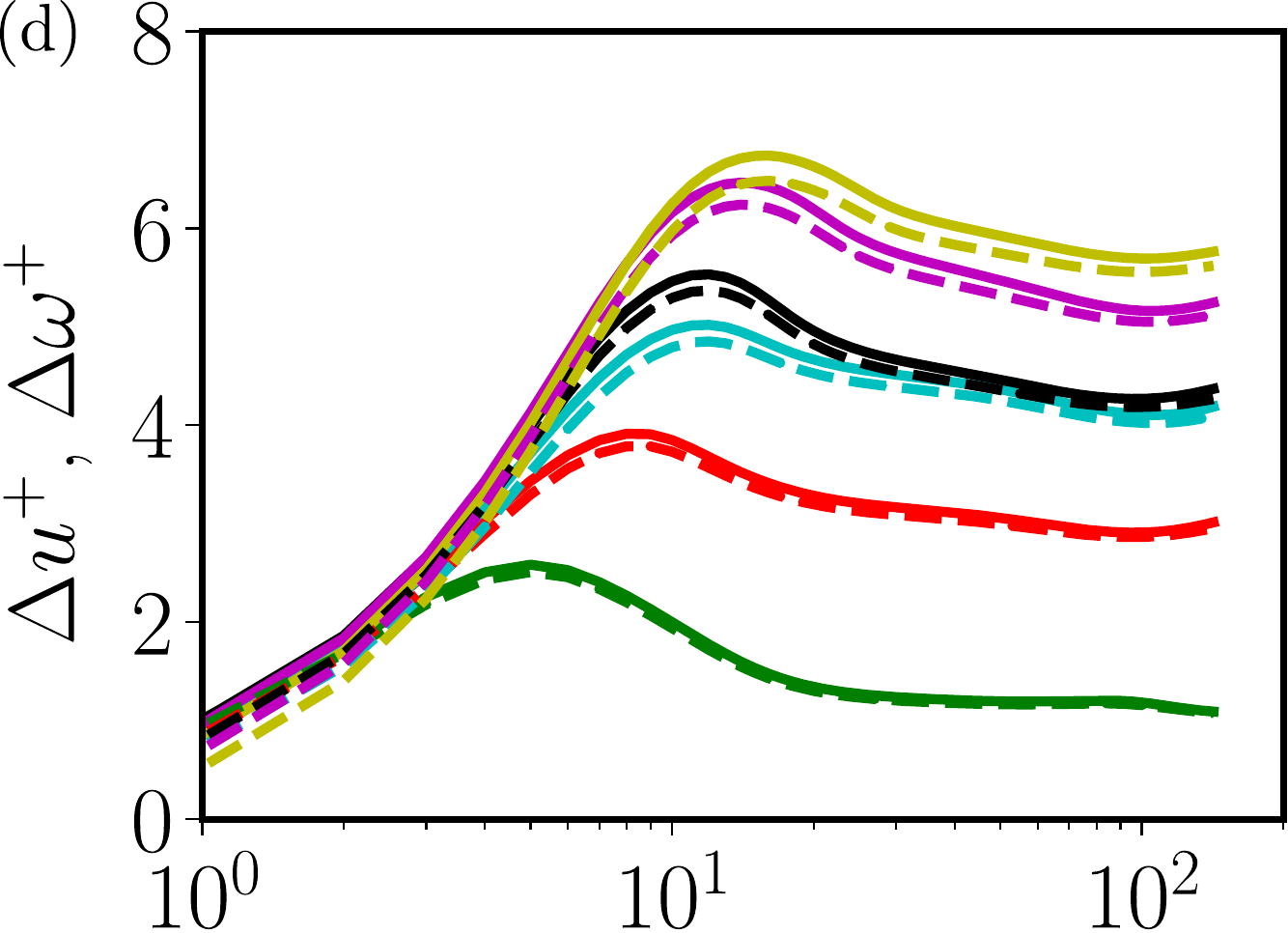}
\end{subfigure}
\begin{subfigure}{.50\textwidth}
  \centering
  \includegraphics[width=0.94\linewidth]{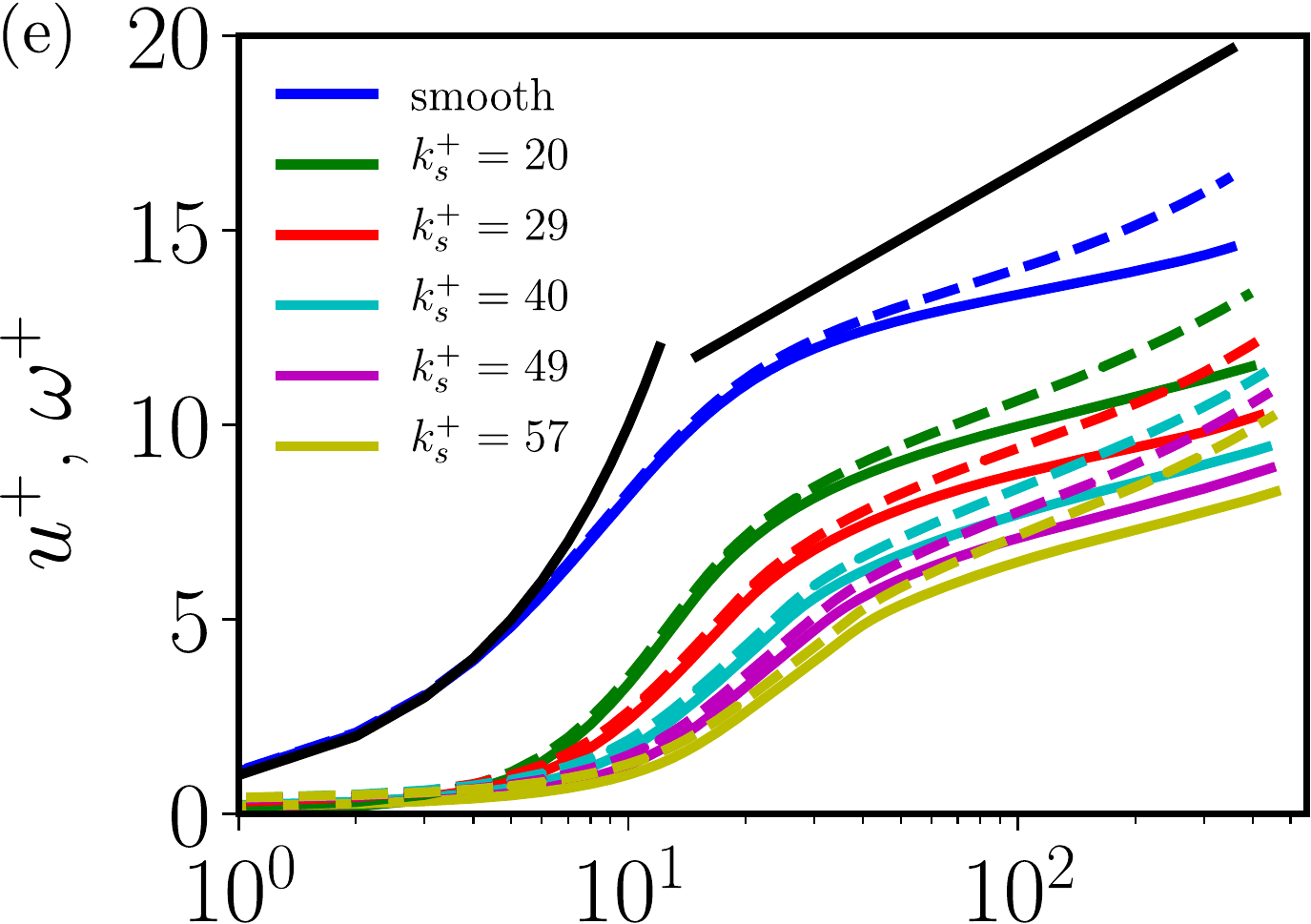}
\end{subfigure}%
\begin{subfigure}{.50\textwidth}
  \centering
  \includegraphics[width=0.94\linewidth]{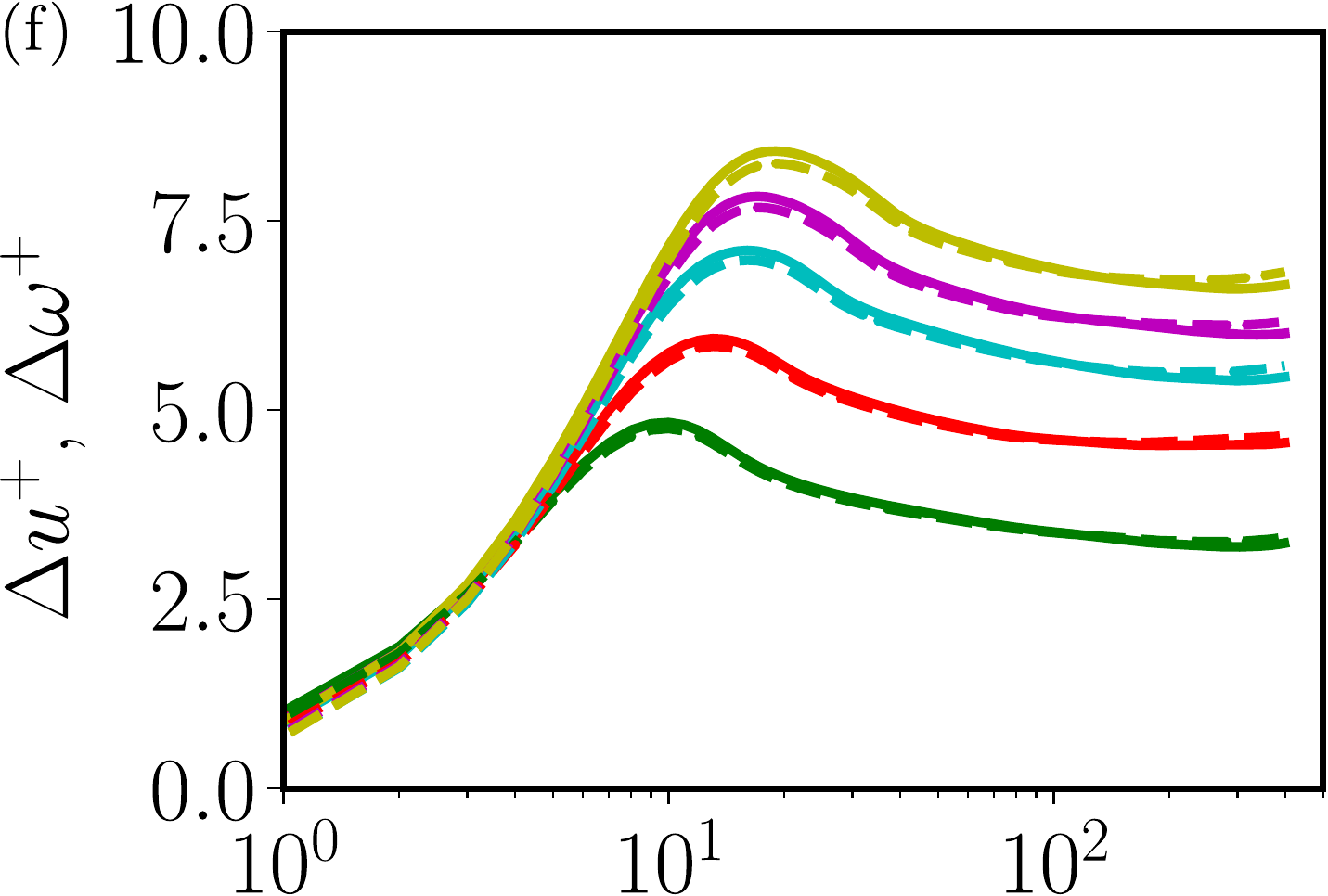}
\end{subfigure}
\begin{subfigure}{.50\textwidth}
  \centering
  \includegraphics[width=0.94\linewidth]{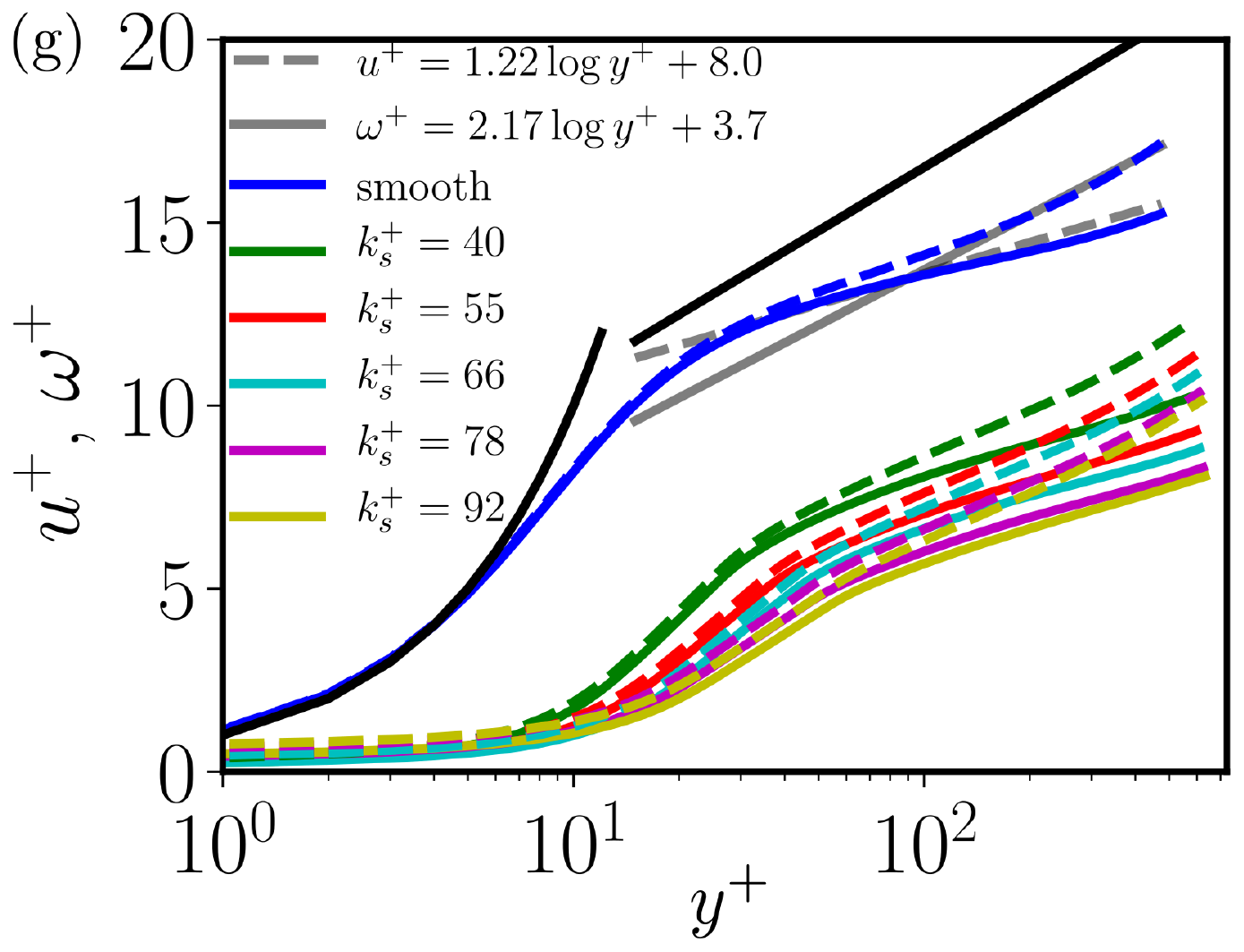}
\end{subfigure}%
\begin{subfigure}{.50\textwidth}
  \centering
  \includegraphics[width=0.94\linewidth]{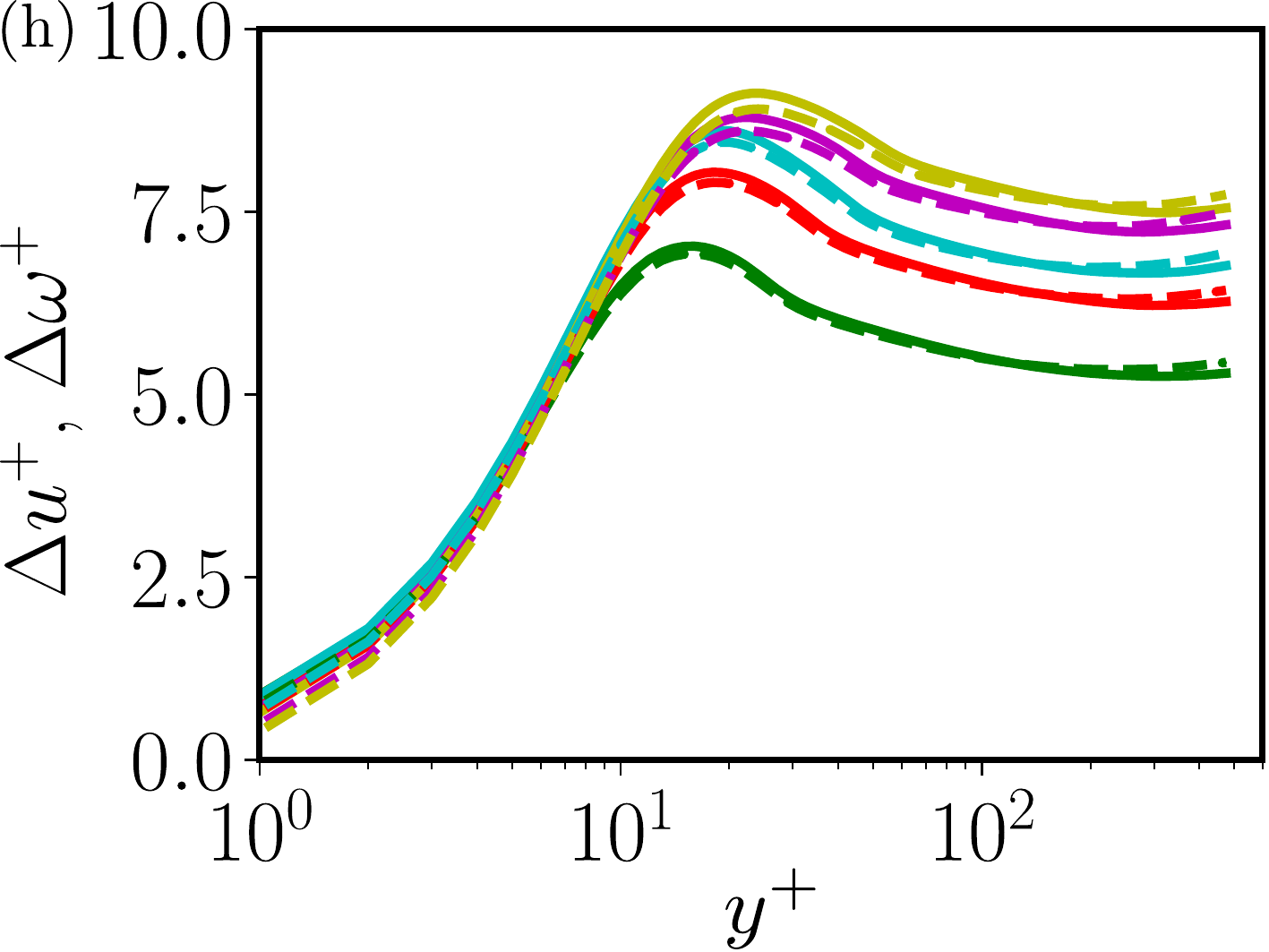}
\end{subfigure}
\caption{($a,c,e,g$) Profiles of the streamwise - azimuthal - velocity $u^+$(\textit{solid}) and the angular velocity $\omega^+$(\textit{dashed}) versus the wall normal distance $y^+$. Black solid lines indicate the viscous sublayer profile $u^+ = y^+$ and the logarithmic law $u^+ = \kappa ^{-1} \log{y^+} + B$, with $\kappa = 0.4$ and $B=5.0$. 
($b,d,f,h$) Profiles of the streamwise velocity shift $\Delta u^+$(\textit{solid}) and the angular velocity shift $\Delta \omega^+$(\textit{dashed}).
Every row corresponds to a constant Taylor number, ($a,b$) $Ta=1.0\times 10^7$, ($c,d$) $Ta=5.0\times 10^7$, ($e,f$) $Ta=5.0\times 10^8$ and ($g,h$) $Ta=1.0\times 10^9$, see Table \ref{table:sim_param}. The grey lines in ($g$) are logarithmic fits to the smooth profiles for $y^+=[150,500]$. }
\label{fig:uprofiles}
\end{figure}

Figure \ref{fig:uprofiles} (\textit{left column)} presents the streamwise (i.e. azimuthal) velocity profiles  $u^+ = \langle (u(r) - u(r_i))\rangle_{z,\theta,t}/u_\tau$ (\textit{solid}) and angular velocity profiles $\omega^+ = \langle u(r_i)-r_iu(r)/r\rangle_{z,\theta,t}/u_\tau$ (\textit{dashed}) versus the wall normal distance $y^+ = r^+ - h_m^+$, where $\langle . \rangle_{z,\theta,t}$ indicates averaging over the streamwise and spanwise directions and in time, and $h_m^+$ is the mean roughness height. Every row corresponds to simulations at constant rotation rate of the inner cylinder (Taylor number), and increasing roughness height. 

In line with the previous observations of \cite{huisman13} and \cite{rodolfo14}, we also find that the logarithmic profiles of the streamwise velocity $u^+$ in smooth-wall Taylor--Couette do not fit the $\kappa = 0.4$ slope, as found in other wall bounded flows (e.g. pipe, boundary layer, channel) - for similar values of the friction Reynolds number $Re_\tau$. However, this asymptotic value is experimentally observed at very high shear rates of $Ta=O(10^{12})$ \citep{huisman13}, much higher than can be obtained by the present DNS. The logarithmic profiles of angular velocity $\omega^+$ have a slope that is closer to the $\kappa = 0.4$ asymptote \citep{rodolfo15}, especially for the higher Ta figure \ref{fig:uprofiles}($g,h$). Here we investigate the effects of roughness on both $u^+$ and $\omega^+$. 
 
For rough wall simulations, the logarithmic region shifts downwards - a hallmark effect of a drag increasing surface. Figure \ref{fig:uprofiles} (\textit{right column}) present the shifts, where $\Delta u^+ = u^+_s - u^+_r$ and $\Delta \omega^+ = \omega^+_s - \omega^+_r$. For lower $Ta$, there is a small but observable difference between $\Delta u^+$ and $\Delta \omega^+$, see figure \ref{fig:uprofiles}($b$), whereas for the higher $Ta$, this difference diminishes, see figure \ref{fig:uprofiles}($f,h$). 

\begin{figure}
\centering
\begin{subfigure}{.50\textwidth}
  \centering
  \includegraphics[width=0.95\linewidth]{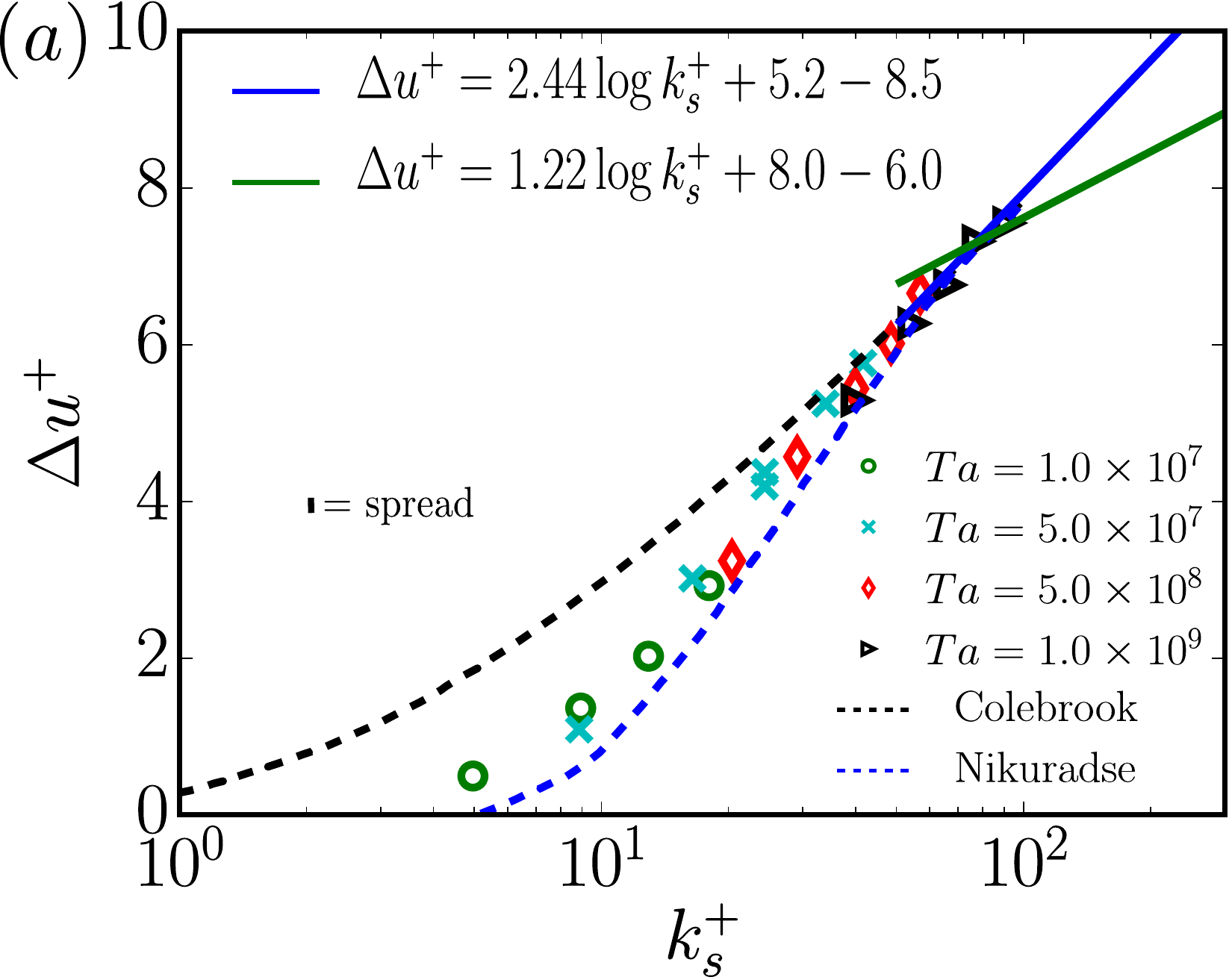}
\end{subfigure}%
\begin{subfigure}{.50\textwidth}
  \centering
  \includegraphics[width=0.95\linewidth]{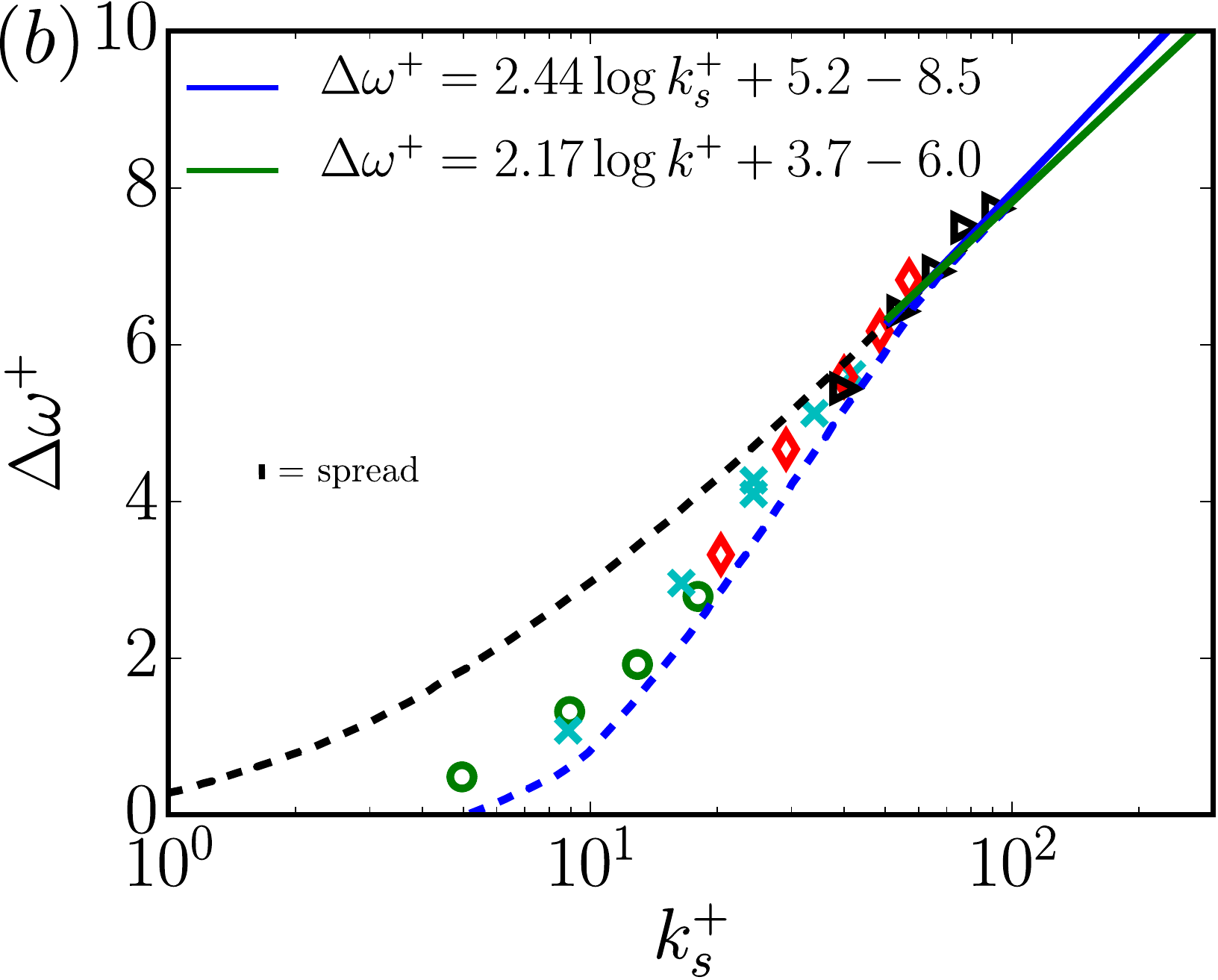}
\end{subfigure}%
\caption{($a$) Azimuthal velocity shift (Hama roughness function) $\Delta u^+$ versus the equivalent sand grain roughness height $k_s^+$, where $k_s^+ = 1.33k^+$. ($b$) Angular velocity shift $\omega^+$ versus $k_s^+$. Close overlap with the Nikuradse curve is observed in the transitionally rough regime.  The overlap is slightly better for the angular velocity shift, for which we also obtain $k_s^+ = 1.33k^+$. The solid blue line represents the fully rough asymptote; $\Delta u^+ = 2.44 \log(k_s^+) + 5.2 -8.5$. The green lines represent the fully rough asymptotes obtained from the simulations, with $\kappa_u,\kappa_\omega, A_u$ and $A_\omega$ extracted from figure \ref{fig:uprofiles}(g). The spread between statistically similar surfaces, with similar mean and maximum heights, is indicated by the vertical bar.  }
\label{fig:roughnessfunc}
\end{figure}
\begin{figure}
\centering
\begin{subfigure}{.50\textwidth}
  \centering
  \includegraphics[width=0.95\linewidth]{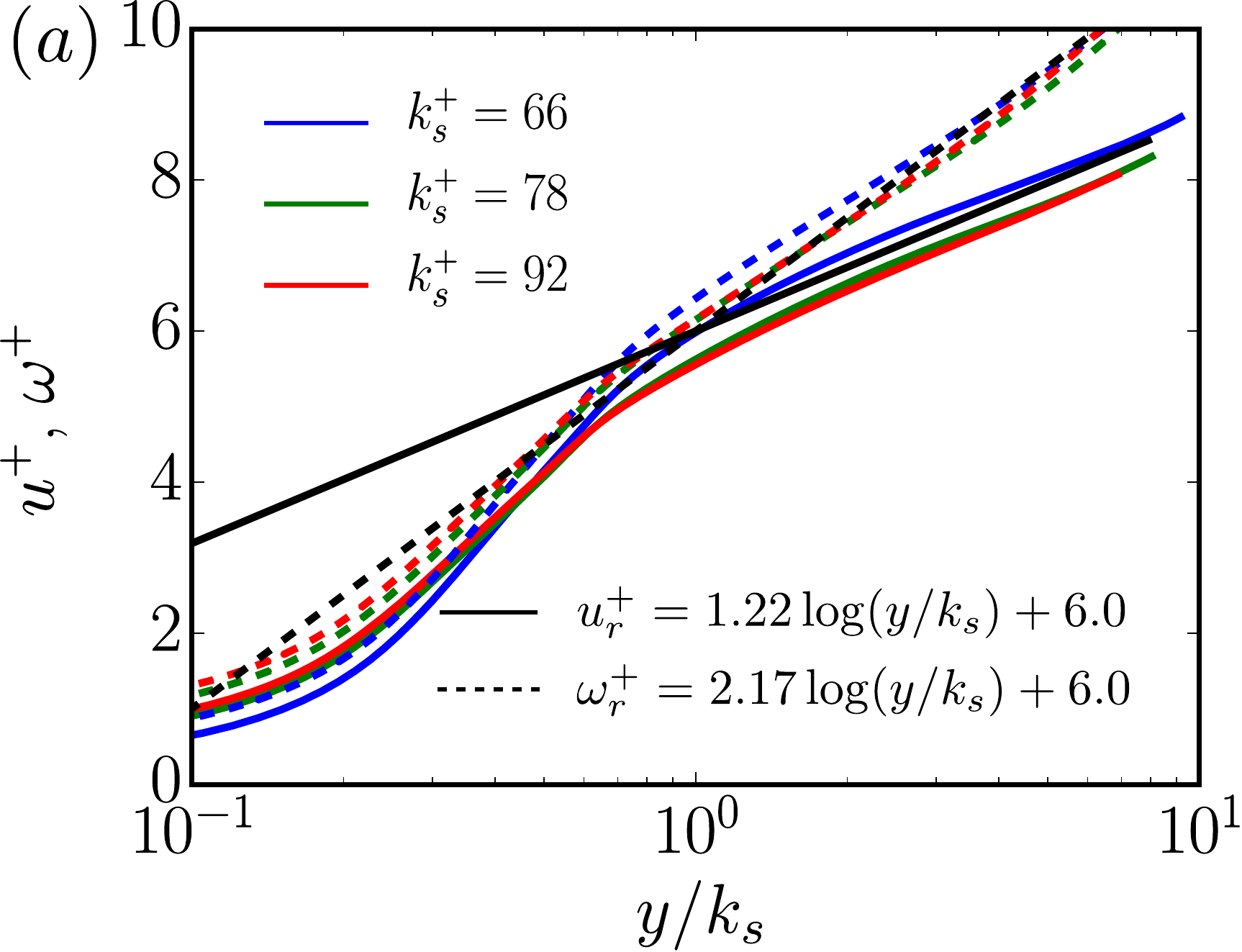}
\end{subfigure}%
\begin{subfigure}{.50\textwidth}
  \centering
  \includegraphics[width=0.95\linewidth]{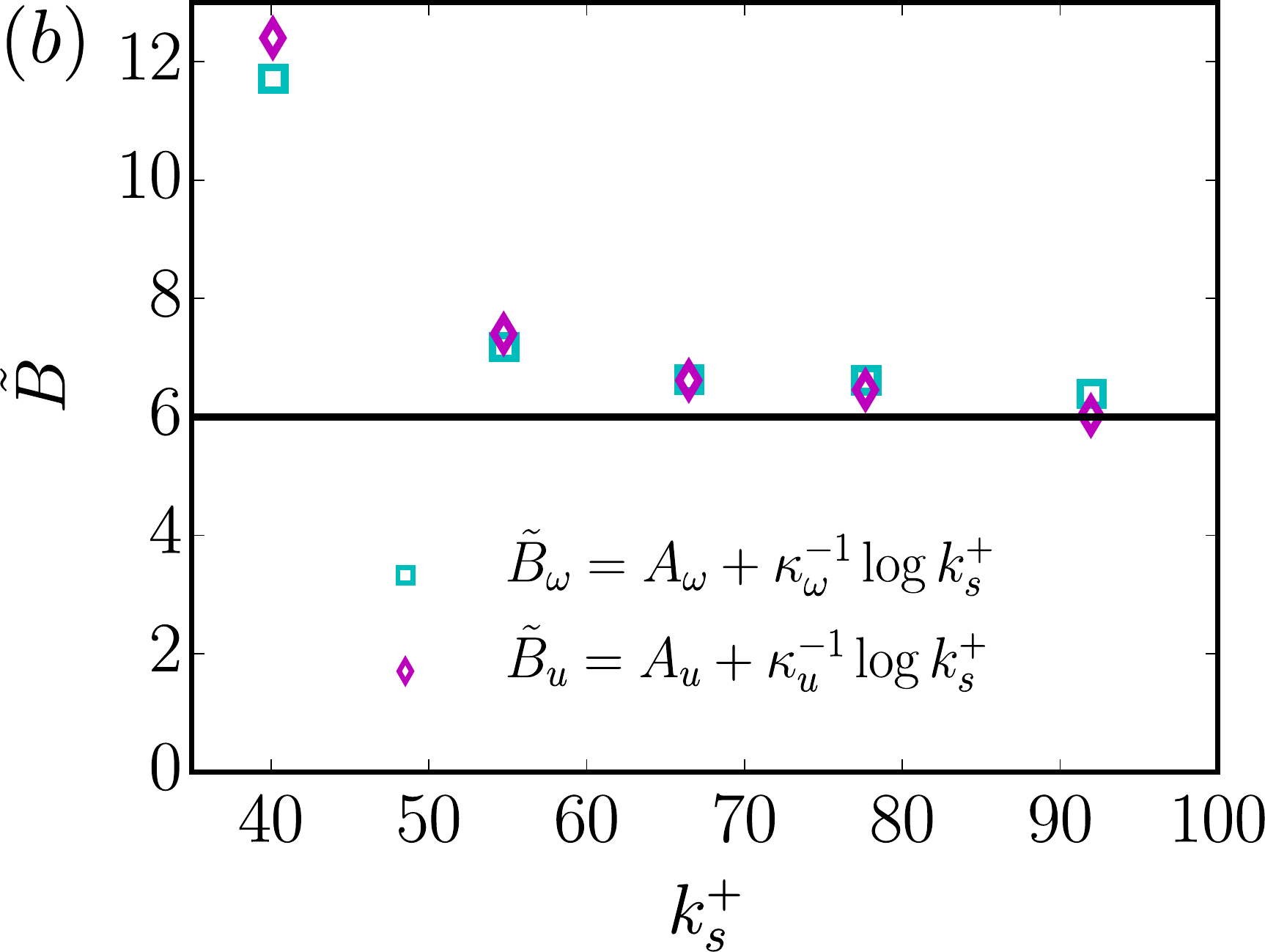}
\end{subfigure}%
\caption{($a$) Azimuthal velocity $u^+$ (\textit{solid}) and the angular velocity $\omega^+$ (\textit{dashed}) versus the wall normal distance $y/k_s$, where $y=(r-h_m)$ and $h_m$ is the mean roughness height. The three simulations with the highest roughness are plotted ($D3$, $D4$ and $D5$ respectively, see Table \ref{table:sim_param}) to convey collapse of the profiles for the fully rough cases. ($b$) The Nikuradse constant $\tilde{B}$ versus the equivalent sand grain roughness height $k_s^+$ for both the azimuthal velocity (\textit{squares}) and the angular velocity (\textit{diamonds}). Horizontal black line at $\tilde{B} = 6.0$ gives the asymptotic value that is observed for fully rough behavior.}
\label{fig:frproof}
\end{figure}
Figure \ref{fig:roughnessfunc}($a,b$) presents the shift of the streamwise and angular velocity profiles, respectively, versus the equivalent roughness height $k_s^+$, for all $Ta$. Care is taken to ensure overlap for varying $Ta$ and similar $k_s^+$, to study the $Ta$ dependence of $\Delta u^+$ and $\Delta \omega^+$. However, despite the varying $Ta$ numbers, all data collapse onto a single curve, with some scatter. Note that scatter is expected due to the randomness of the surfaces, which are reproduced for every simulation. To obtain an estimate of the expected scatter, we run two simulations with statistically similar surfaces. These are indicated by B3 and BY in table \ref{table:sim_param} and the velocity profiles are found in figures \ref{fig:uprofiles}($c,d$). We find a difference between the two cases of $\lesssim 0.2 \Delta u^+, 0.2 \Delta \omega^+$. 
 
A comparison with the findings of \cite{Nikuradse33} can be carried out by scaling the fully rough regime to obtain $k_s^+ = Ck^+$, where $C$ is a constant that depends on the surface topology. In figure \ref{fig:frproof}($a$) we plot the velocity profiles versus $(r-h_m)/k_s$ for the highest roughness (D3,D4 and D5 respectively, see Table \ref{table:sim_param}). Excellent collapse of the D4 and D5 profiles indicates that those simulations are indeed fully rough. In this fully rough regime viscosity can be neglected ($y\gg k \gg \delta_\nu$), whereas the velocity profile is also independent of the system outer length scales ($y \ll d$) i.e. the \textit{overlap argument} \citep{pope2000}. The gradient of the velocity profile becomes; $\frac{d\langle U \rangle}{dy} = \frac{u_\tau}{y}\Phi(y/k)$,  where $\Phi(y/k)$ is a universal function that will go to $1/\kappa$. Integration then gives; 
\begin{equation}
u_r^+ = \frac{1}{\kappa}\log{(y/k)} + B = \frac{1}{\kappa} \log{(y/k_s)} + \tilde{B}
\label{eq:overlap}
\end{equation}
where $B$ is a constant and $y=r-h_m$. $\tilde{B}$ is the Nikuradse constant. 
The roughness function in the fully rough regime, (i.e. the fully rough asymptote), is obtained by subtracting (\ref{eq:overlap}) from the smooth wall profile 
$u_s^+ = 1/\kappa \log{(y^+)} + A$ and rescaling it to overlap with the Nikuradse data:
\begin{equation}
\Delta u^+ = \frac{1}{\kappa}\log{(k_s^+)} + A - \tilde{B}.
\label{eq:frasymptote}
\end{equation}
In figure \ref{fig:roughnessfunc}($a$), the blue solid line is the fully rough asymptote, with $\kappa,A,\tilde{B}$ as found in pipe flow \citep{pope2000}. The green solid line is the fully rough asymptote as obtained from our simulations. $\kappa_u^{-1} = 1.22$ and $A_u = 8.0$ are taken from the fit of the smooth wall simulation at identical $Ta$ as the fully rough cases (figure \ref{fig:uprofiles}$g$). The fits are in the domain $y^+=[150,500]$, as there the slope becomes approximately constant (figure \ref{fig:uprofiles}$h$). $\tilde{B}$ is plotted in figure \ref{fig:frproof}($b$), where we find that $\tilde{B}\approx6.0$ for the fully rough cases. The mismatch of the slopes in the fully rough regime makes a rescaling to find $k_s^+$ \textit{a priori} impossible - a statement that we wish to emphasize. However, to proceed with the comparison of the transitionally rough cases in TC and pipe flow, we choose to rescale the fully rough cases (D4 and D5) with the Nikuradse fully rough asymptote in  figure \ref{fig:roughnessfunc}. We find that $k_s^+ = 1.33 k^+$ and very close collapse of our data with the Nikuradse data. 

In parallel, we analyse the behavior of $\Delta \omega^+$ versus $k_s^+$, shown in figure \ref{fig:roughnessfunc}($b$). Again, the blue solid line represents the fully rough asymptote of Nikuradse. The green solid line is the fully rough asymptote obtained from fits ($y^+=[150,500]$) of the smooth wall angular velocity profile at identical $Ta$ as the fully rough cases, see figure \ref{fig:uprofiles}($g$). We find $\kappa_\omega^{-1} = 2.17$ ($A_\omega=3.7$), close to the asymptotic value $\kappa^{-1}=2.44$. Although the differences are marginally, $\Delta \omega^+$ fits to the Nikuradse data slightly better than $\Delta u^+$ (note that also here the rescaling is with the Nikuradse data, $k_s=1.33k^+$). However, the major difference is the closeness of the fully rough asymptotes. 

These results suggest that the near-wall effects of transitionally rough sand grains (and other rough surfaces) in TC flow are similar to the effects of transitionally rough sand grains (and other rough surfaces) in pipe flow (and other canonical systems). Further, we find that these transitionally rough effects are independent of slope of the logarithmic law, whereas in the fully rough regime, they, \textit{a priori}, depend on this slope. This is confirmed with similar values of $\Delta u^+$ at similar $k_s^+$, for varying $Ta$, see figure \ref{fig:roughnessfunc}. Also the similarity between $\Delta u^+$ and $\Delta \omega^+$ in the transitionally rough regime confirms this, whereas the fully rough asymptotes are very dissimilar. We like to point out that simulations (or experiments) at high enough $Ta$ ($=10^{12}$ \citep{huisman13}), where $\kappa = 0.4$, then are expected to also give a collapse to the Nikuradse data in the fully rough regime.

\subsection{Global response}\label{global}
\begin{figure}
\centering
\begin{subfigure}{.5\textwidth}
  \centering
  \includegraphics[width=1.0\linewidth]{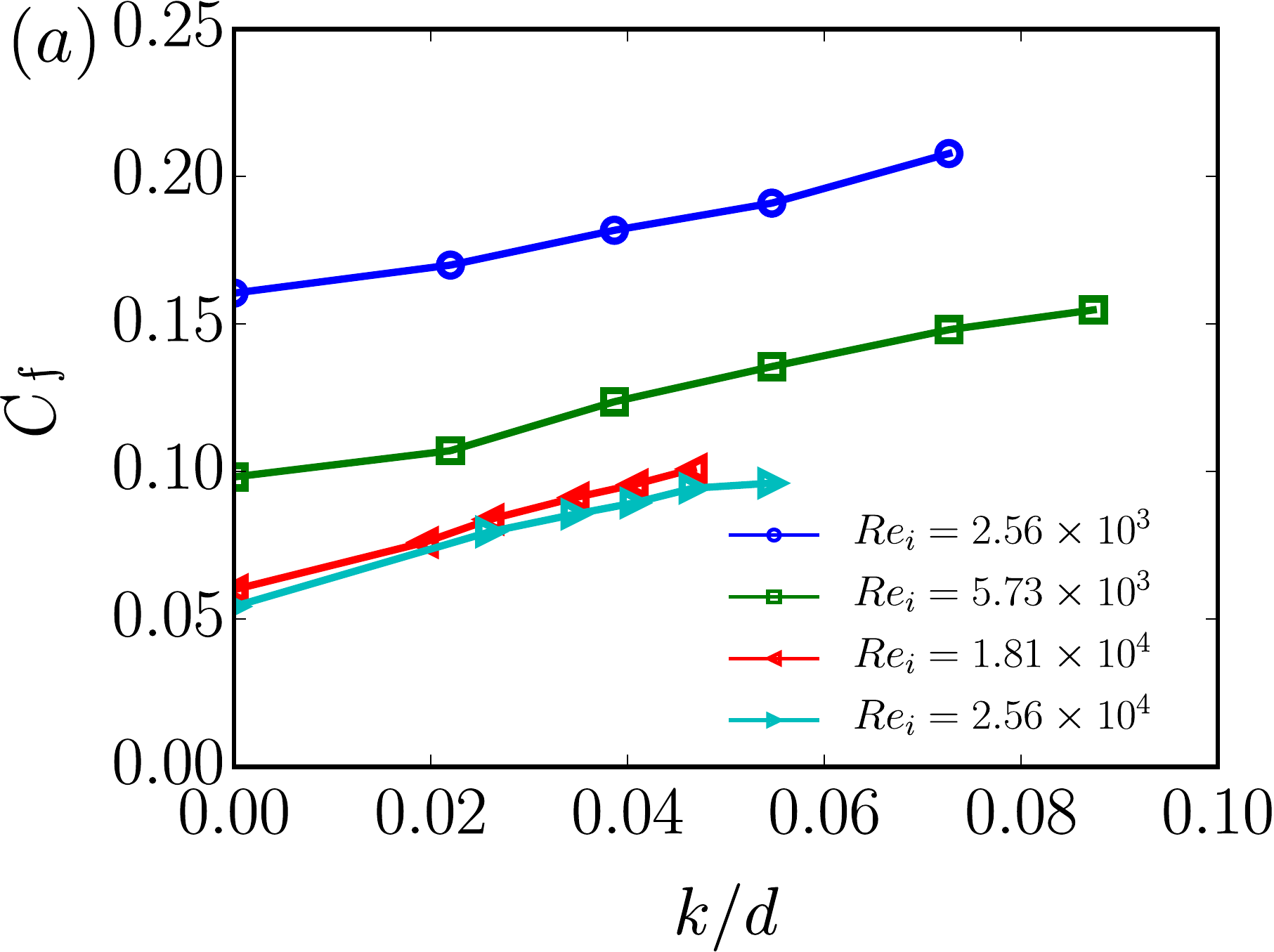}
\end{subfigure}%
\begin{subfigure}{.5\textwidth}
  \centering
  \includegraphics[width=1.0\linewidth]{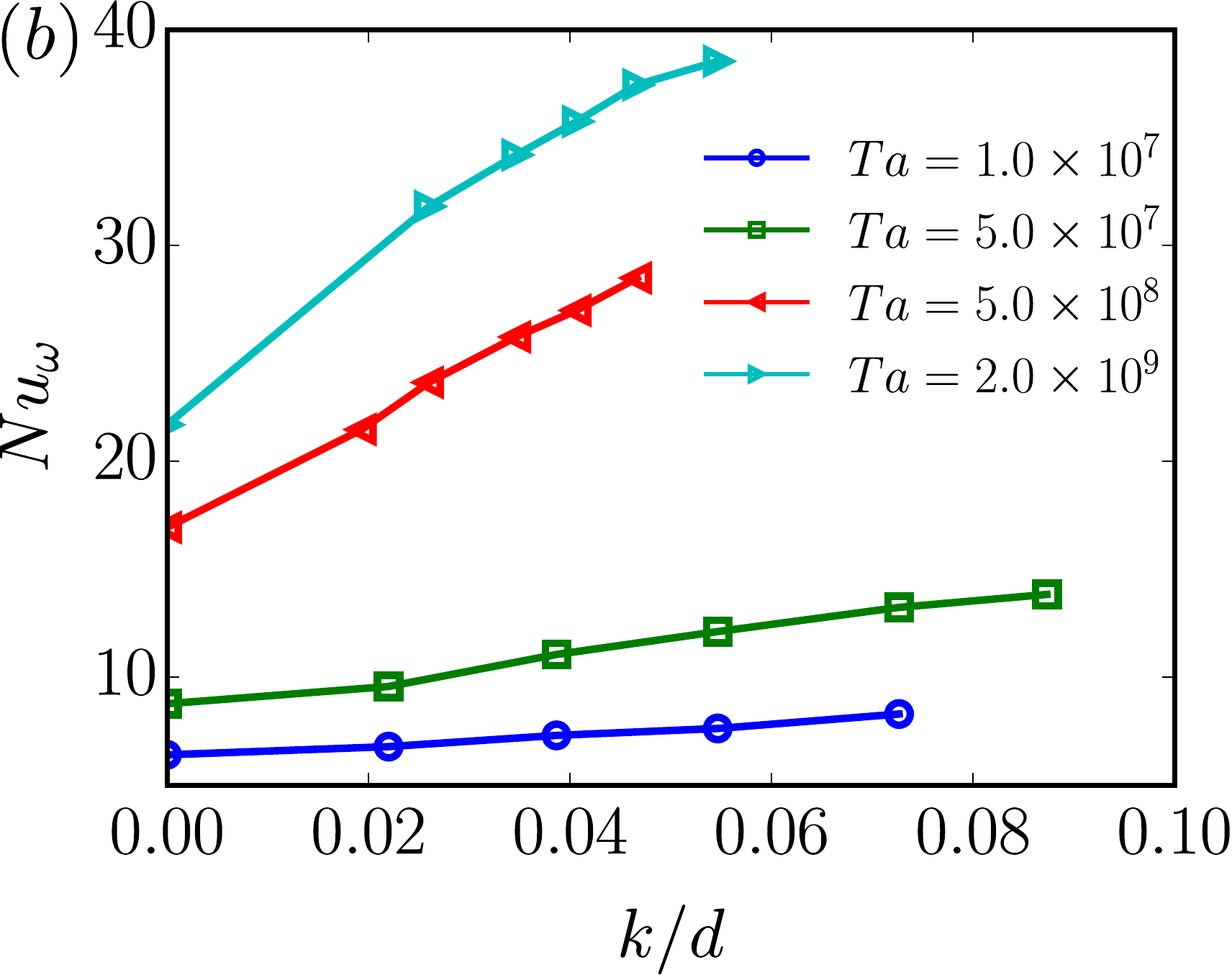}
\end{subfigure}
\caption{Profiles of $(a)$ the friction factor $C_f$, and $(b)$ the Nusselt number $Nu_\omega$ versus the roughness height. $k/d$ is the roughness height $k$ relative to the gap width $d$.  }
\label{fig:nuheight}
\end{figure}
Figure \ref{fig:nuheight}($a$) presents the friction factor $C_f$ versus the dimensionless roughness height $k/d$ for varying $Re_i$. For increasing $Ta$ numbers, the friction factor decreases (e.g. figure 7.23 in \cite{pope2000}), an indication that pressure is not yet dominant. For constant $Ta$, as expected, the friction factor increases for increasing roughness height. Figure \ref{fig:nuheight}($b$) then presents the global response in terms of $Nu_\omega$. We observe an increase in $Nu_\omega$ for increasing $Ta$, corresponding to the increased transport of the angular velocity that is due to the increased turbulent mixing. Higher roughness leads to increased $J^\omega$ as compared to the smooth wall at the same $Ta$, which also relates to a higher intensity of the turbulent mixing (the $r^3(\langle u_r\omega\rangle_{z,\theta,t})$ term of equation (\ref{eq:jom})) and more plumes ejecting from the boundary layer radially outwards \citep{zhu17}, on which we will elaborate in \S \ref{local}. 
\begin{figure}
\centering
\begin{subfigure}{.5\textwidth}
  \centering
  \includegraphics[width=1.0\linewidth]{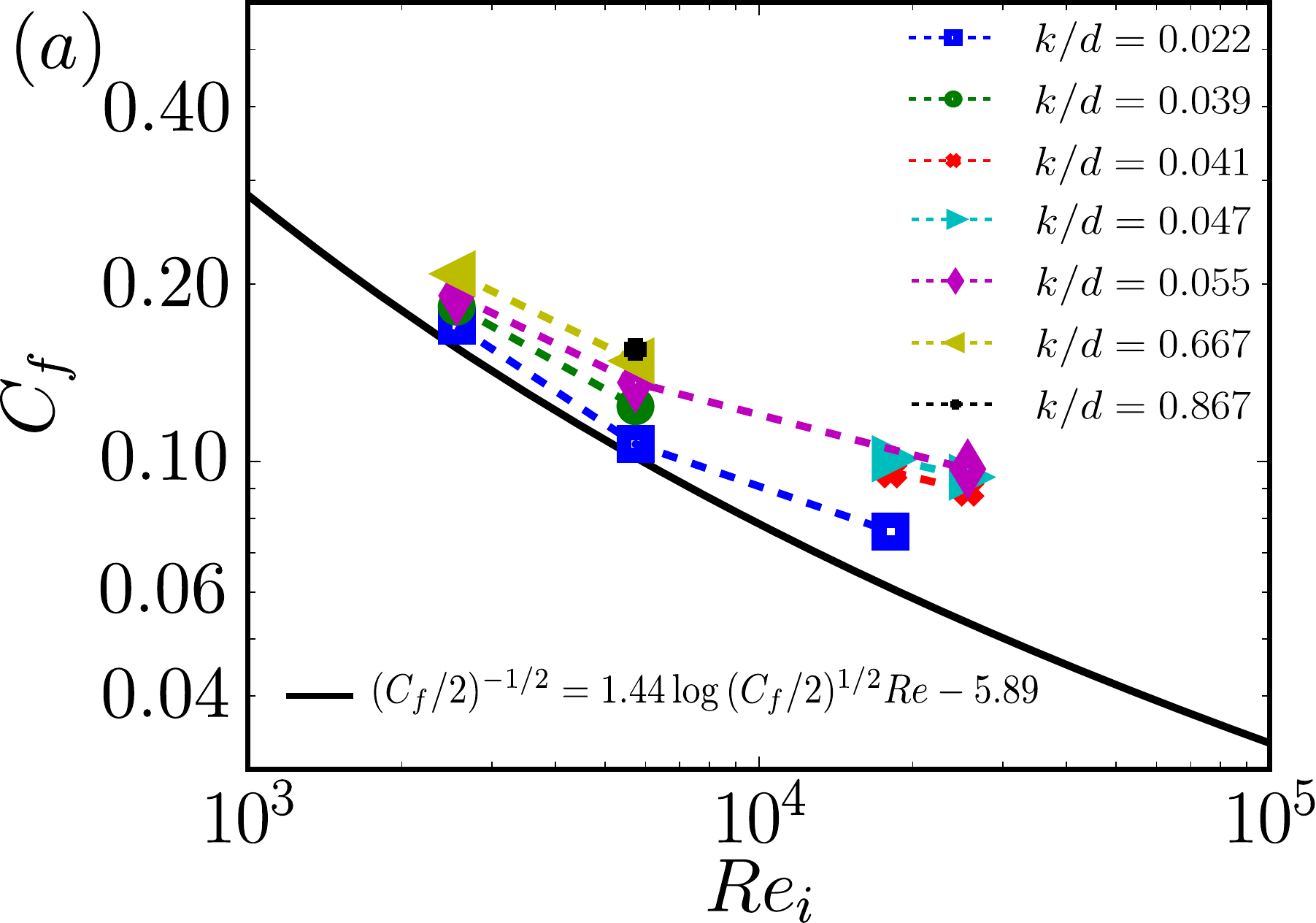}
\end{subfigure}%
\begin{subfigure}{.5\textwidth}
  \centering
  \includegraphics[width=1.0\linewidth]{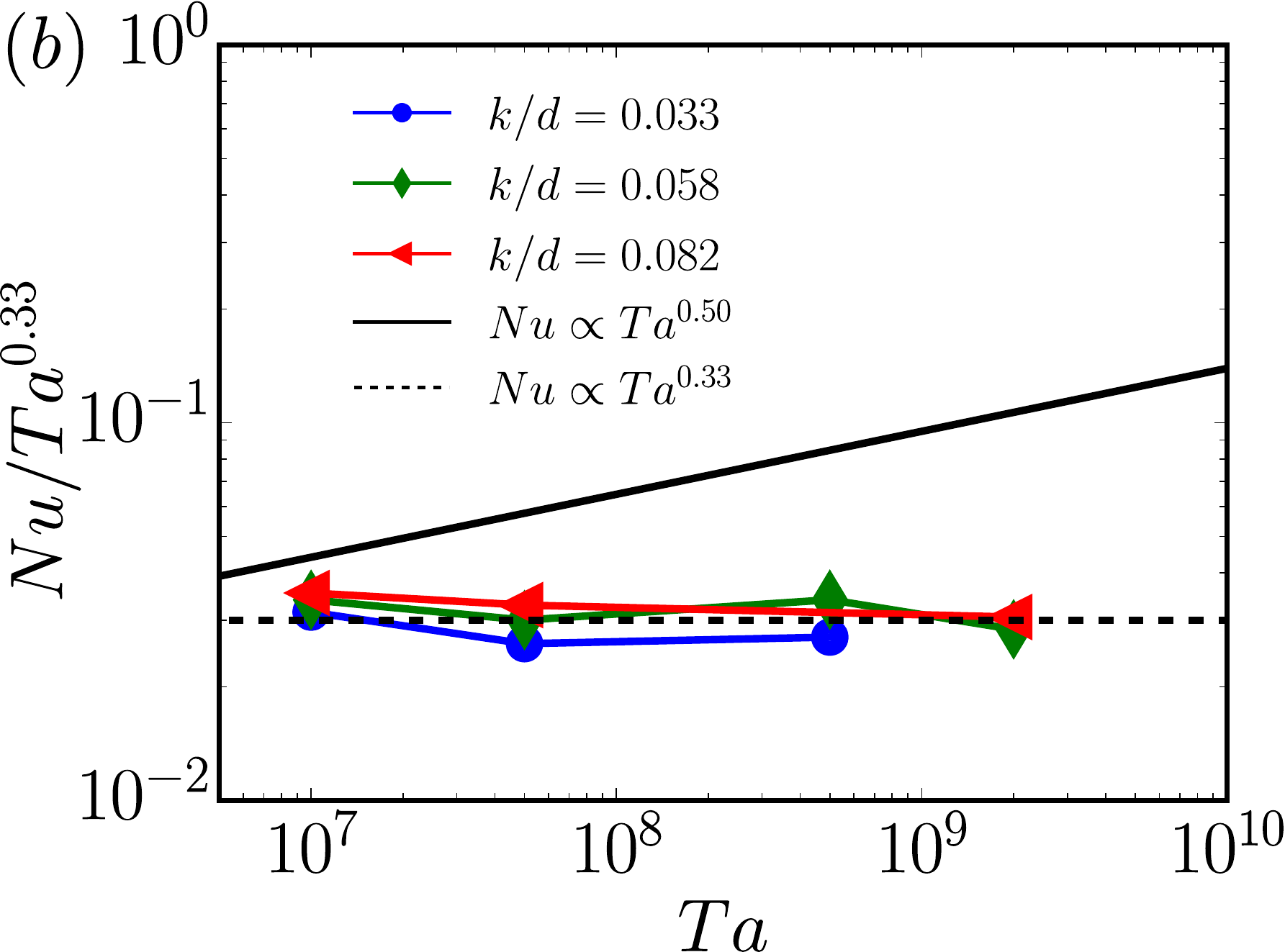}
\end{subfigure}
\caption{$(a)$ Moody representation, showing the friction factor $C_f$ as a function of the inner cylinder Reynolds number $Re_i$  for varying roughness height $k/d$. The solid line is the fit of the Prandtl friction law to the smooth wall simulation data. $(b)$ Compensated plot of the Nusselt number versus the Taylor number for constant $k/d$. In this regime, $Nu\propto Ta^{0.33}$. The solid black line indicates the assymptotic scaling of $Nu\propto Ta^{0.50}$. }
\label{fig:prandtl}
\end{figure}
By assuming a logarithmic profile, and integrating this profile over the entire gap (thereby neglecting the contributions of the viscous sublayer), we arrive at an implicit equation for the friction factor $C_f$, namely the celebrated Prandtl's friction law:
\begin{equation}
(C_f/2)^{-1/2} = C_1\log((C_f/2)^{1/2}Re_i) + C_2.
\end{equation}
Figure \ref{fig:prandtl} shows the friction factor $C_f$ versus the inner cylinder Reynolds number $Re_i$, for both smooth wall as the rough wall data. An upward shift of the friction factor for increasing roughness height is consistent with  what is observed for sand grains in pipe flow  \citep{Nikuradse33} and recently also for tranverse ribs in Taylor--Couette flow \citep{zhu18}.
Note that this upward shift, is directly related to the downward shift of the mean streamwise velocity profile (the roughness function). Since the friction factor and the Nusselt number are related, as expressed in equation (\ref{eq:cf}), we expect the Nusselt number to increase, for increasing roughness height. This is confirmed in figure \ref{fig:prandtl}($b$), where we plot the Nu number versus the Ta number. The number of simulations with constant $k/d$ is limited, and we vary the Ta number over $2$ decades only. However, we clearly observe that the asymptotic, ultimate scaling of $Nu_\omega\propto Ta^{0.5}$, as found for fully rough transverse ribs in \citep{zhu18}, is not reached. This is expected,
as for Ta numbers the system is still in the transitionally rough TC regime, where vicous forces still strongly contribute to $Nu_\omega$, and only the inner cylinder is covered with roughness.

\subsection{Flow structures}\label{local}
\begin{figure}
\captionsetup[subfigure]{labelformat=empty}
\centering
\begin{subfigure}{1\textwidth}
  \centering
  \includegraphics[width=1\linewidth]{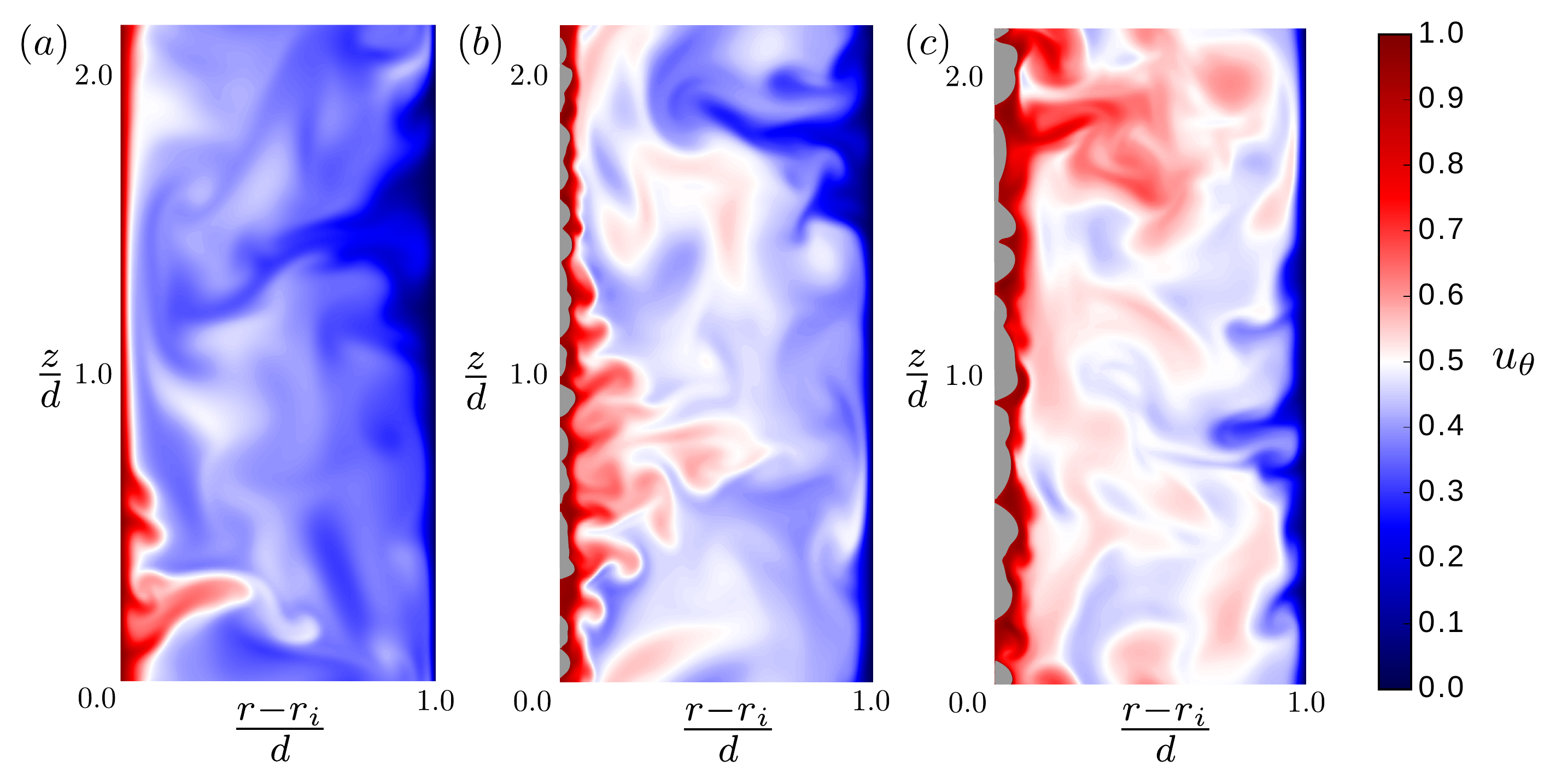}
  \caption{\textbf{Classical turbulent state} Contour fields of the instanteneous azimutal velocity profile $u_\theta(r,\theta,z,t)$ for $Ta=5.0\times 10^7$ in the meridional plane. ($a$) Smooth wall simulation (BS) in which we observe one ejecting plume and one impacting plume. ($b$) Rough inner cylinder $k/d=0.039$ (B2), with the roughness indicated in grey, exhibiting more plumes ejecting from the inner cylinder radially outwards. ($c$) Rough inner cylinder $k/d=0.073$, with the roughness indicated in grey, (B4) leading to a more chaotic flow field, with enhanced mixing and enhanced radial transport of the conserved angular velocity flux.}
\end{subfigure}
\begin{subfigure}{1\textwidth}
  \centering
  \includegraphics[width=1\linewidth]{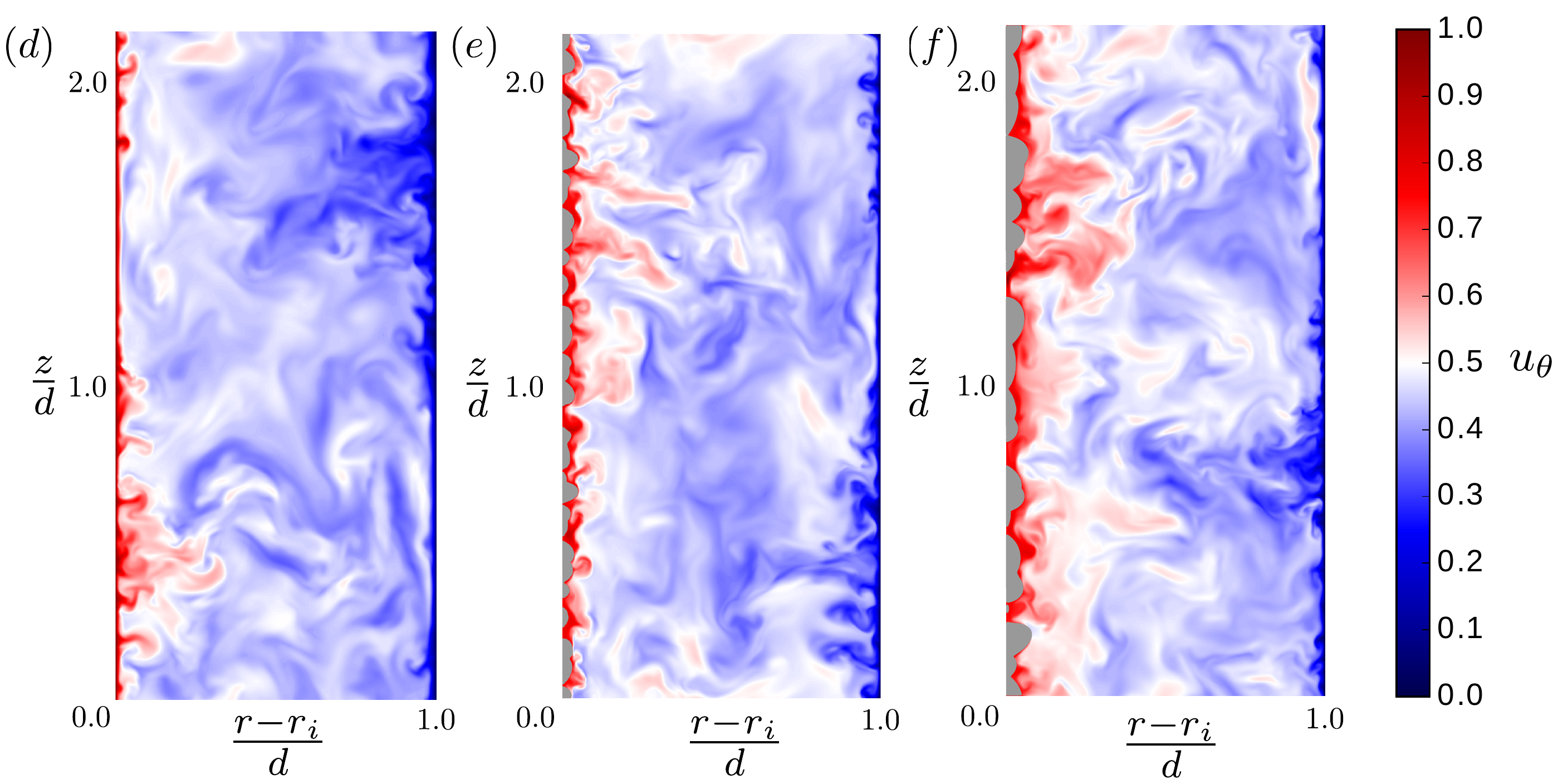}
\caption{\textbf{Ultimate turbulent state} Contour fields of the instanteneous azimutal velocity profile $u_\theta(r,\theta,z,t)$ for $Ta=1.0\times 10^9$ in the meridional plane. ($d$) Smooth inner cylinder (DS) with many plumes ejecting, considerably more chaotic than in (a). ($e$) Inner cylinder wall roughness (indicated in grey) $k/d = 0.035$ (D2) and ($f$) inner wall roughness $k/d = 0.055$ (D5). For the rough cases, we observe more plumes ejecting and more mixing in the bulk, leading to enhanced radial transport of the angular velocity $J^\omega$, expressed in a higher $Nu_\omega$.}
\end{subfigure}
\caption{}
\label{fig:snapshots}
\end{figure}
To obtain a qualitative understanding of the effect of inner cylinder roughness on the turbulent flow in the gap, we present two series of snapshots of the streamwise azimuthal velocity $u_\theta(r,\theta,z)$. Figure  \ref{fig:snapshots}($a-c$) exhibit the snapshots for $Ta=5.0\times10^7$. It is known, and observed here, that for this Taylor number the coherence length of the dominant flow structures becomes smaller than the gap width $d$, and turbulence develops in the bulk \citep{grossmann16}. On the other hand, the boundary layers remain predominantly laminar and as such the regime is referred to as the `classical regime' of TC turbulence. A divergent colormap is chosen to highlight the turbulent structures in the bulk. A snapshot for the smooth inner cylinder simulation is presented in figure \ref{fig:snapshots}($a$). At $z/d\approx0.3$, one observes an ejecting structure (plume) that detaches from the inner cylinder laminar boundary layer at the location of an adverse pressure gradient. Locally, where this ejecting plume detaches, the flow will be different (i.e. more chaotic) to that in the other parts of the boundary layer. As such, one also expects the local variables (e.g. skin friction, turbulence intensity) to be different. Later we will therefore also employ local averaging, to investigate the spatial differences in the flow associated with this structure (\cite{zhu18prl,vdpoel15prl,rodolfo16}). The ejecting and impacting (located at $z/d\approx1.3$) plumes have very strong radial velocity components $u_r$. From the first term on the right-hand side of (\ref{eq:jom}), $r^3(\langle u_r\omega\rangle_{z,\theta,t})$, we then directly see that they strongly contribute to $Nu_\omega$. This brings us to the remaining (figure \ref{fig:snapshots} $b$ and $c$) snapshots. Many more, small, plumes are seen to eject from the inner cylinder. The roughness there promotes the detachment of ejecting structures and in that way contributes to a higher $Nu_\omega$. An increase in the level of turbulence, as suggested by the increased level of turbulence dissipation, is quantitatively reflected by a decrease in the Kolmogrov scale ($\eta = (\nu^3/\epsilon)^{1/4}$), namely $\eta/d=7.1\times10^{-3}$ for the smooth wall case BS and $\eta/d=6.5\times10^{-3}$ for the highest roughness case B5. Note that the decrease in the Kolmogorov scale $\eta$ is only small, since $\eta/d \propto (\epsilon d^4 /\nu^3)^{-1/4}$. For TC flow, the volume averaged dissipation rate $\epsilon$ is related to the angular velocity transport $Nu_\omega$  with: $\epsilon = \nu^3 d^{-4} \sigma^{-2} (Nu_\omega - 1)Ta + \epsilon_{lam}$, where $\epsilon_{lam}$ is the laminar volume averaged dissipation rate, $d$ is the gap width of the setup and $\sigma$ is a geometric parameter \citep{eckhardt07}. As such, we see that $\eta/d \propto Nu_\omega^{-1/4}$ only. 

Figure \ref{fig:snapshots}($d-f$) presents snapshots of a flow in the ultimate turbulent state at $Ta=1.0\times 10^9$ \citep{grossmann16}. Although less pronounced than for $Ta=5.0\times 10^7$, we still observe distinct ejecting and impacting regions, indicating the survival of the turbulent Taylor rolls. A similar rationale as applied above, to the classical turbulence case, can also be used to explain the enhancement of the Nusselt number for rough inner cylinders in the ultimate turbulent state. In fact, we can also observe more intense plumes for the highest roughness (D5, figure \ref{fig:snapshots}$f$), in comparison to a lower roughness case (D2, figure \ref{fig:snapshots}$e$). Note that here we do not observe the stable vortex formation in between roughness elements and the associated ejection of plumes from sharp peaks, as was reported by \cite{zhu16} for grooved cylinders, for similar Taylor numbers and roughness heights. The increase in the turbulence level is also quantitatively confirmed by a decrease in the Kolmogorov scale here, $\eta/d=2.7\times10^{-3}$ for the smooth wall case DS and $\eta/d=2.1\times10^{-3}$ for the highest roughness case D5.

\subsection{Roughness sublayer}\label{sub}
\begin{figure}
\centering
\hspace*{-7mm}
\includegraphics[width=1.0\linewidth]{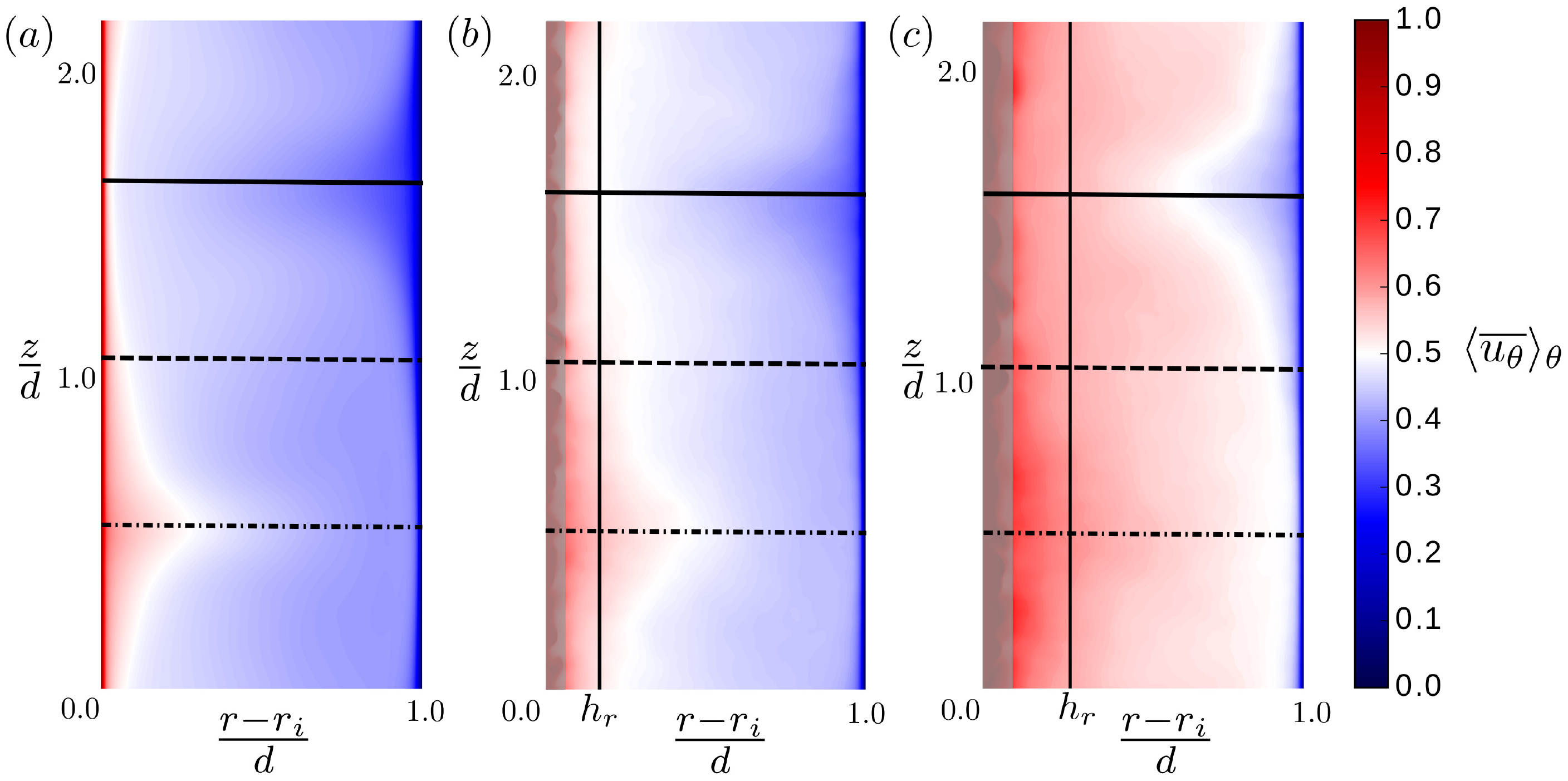}
\caption{Contour field of the mean azimuthal velocity $\langle \overline{u_\theta} \rangle _{\theta}$ in the meridional plane for $Ta=1.0\times 10^9$. ($a$) Smooth inner cylinder (DS) ($b$) Inner cylinder wall roughness $k/d = 0.0517$ (D2) and ($c$) inner wall roughness $k/d = 0.0818$ (D5). The solid vertical lines indicate the height of the roughness sublayer $h_r$, calculated over the entire cylinder height. \protect\tikz[baseline]{\protect\draw[line width=0.5mm,dash dot] (0,.8ex)--++(1,0) ;}~ plume ejection regions, \protect\tikz[baseline]{\protect\draw[line width=0.5mm, densely dashed] (0,.8ex)--++(1,0) ;}~ sheared regions, \protect\tikz[baseline]{\protect\draw[line width=0.5mm] (0,.8ex)--++(1,0) ;}~ plume impacting regions.}	
\label{fig:sublayeru}
\end{figure}
The existence of Taylor roll structures is already anticipated in the snapshots of the instantaneous flow in figure \ref{fig:snapshots}, from which we observe the ejecting and impacting plume regions. Contour plots of the time and azimuthally averaged azimuthal velocity field, as presented in figure \ref{fig:sublayeru}, confirm this. Note that the Taylor roll is spatially fixed, allowing for convenient averaging over impacting (solid line), shearing (dashed line) and ejecting (dashed dotted line) regions, a method that we also have employed in RB flow \citep{vdpoel15prl}. For an increasing roughness height, the white region (representing $\langle \overline{u_\theta}\rangle_\theta \approx 0.5$) shifts radially outwards and the azimuthal velocity in the bulk increases. This process previously has been seen in \cite{zhu18}, where it is referred to as the bulk velocity `getting slaved to' to the velocity of a cylinder covered with roughness, reflecting the enhanced drag on the rough side.
\begin{figure}
\centering
\hspace*{-7mm}
\includegraphics[width=1.0\linewidth]{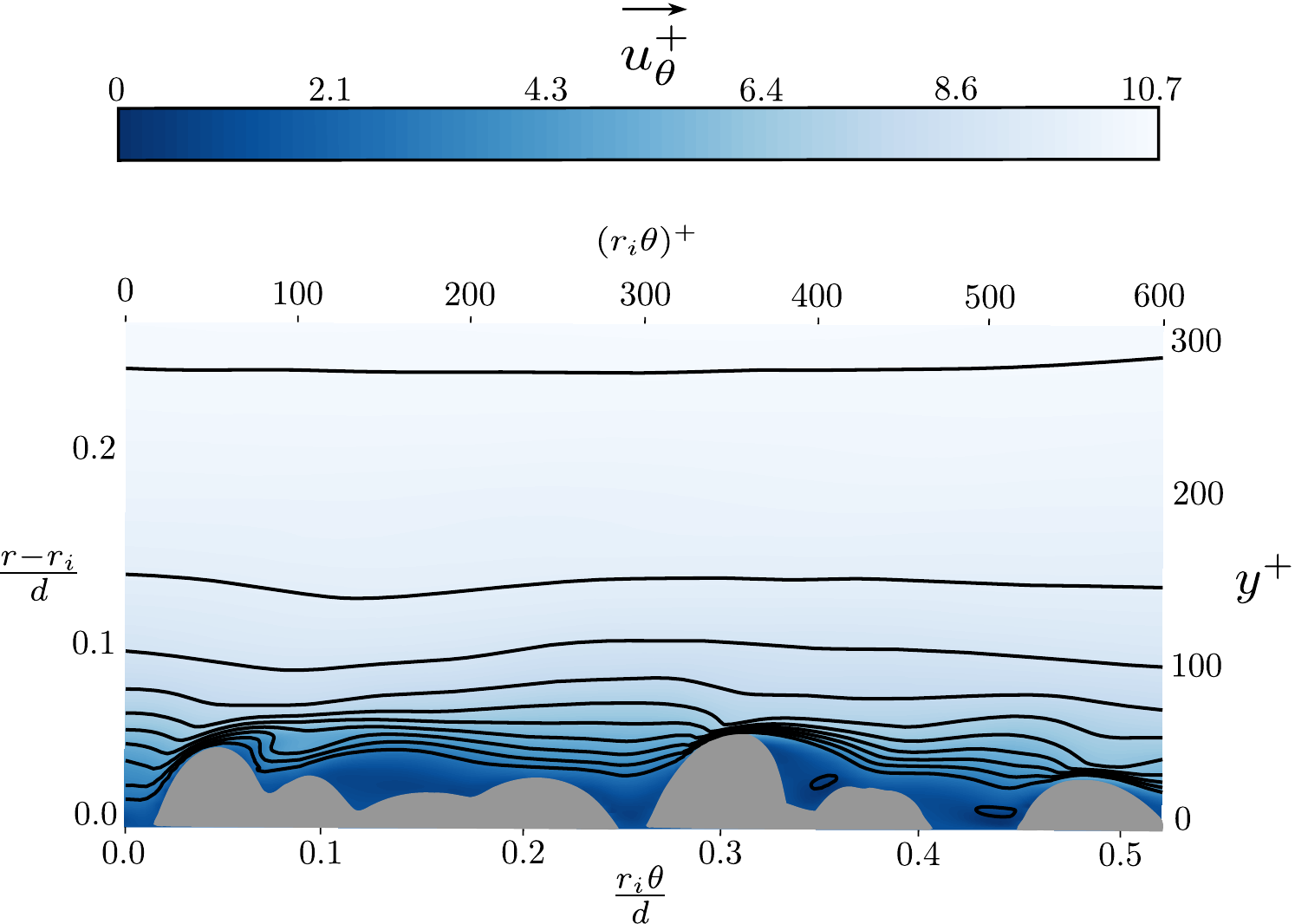}
\caption{Contour of the time-averaged azimuthal velocity $u_{\theta}^+$, zoomed in on only a few roughness elements, indicated in grey, for $Ta=1.0\times10^9$ (D2). Isolines of $u_{\theta}^+$ overlay the contour. On the vertical axis, we display the wall normal coordinate normalized by the gap width $d$ (\textit{left}) and in viscous units (\textit{right}). On the horizontal axis, we display the azimuthal coordinate, normalized by the gap width $d$ (\textit{below}) and in viscous units (\textit{above}). The arrow indicates the direction of the mean flow.}
\label{fig:sublayeru2}
\end{figure}
\begin{figure}
\centering
\begin{subfigure}{.5\textwidth}
  \centering
  \includegraphics[width=1.0\linewidth]{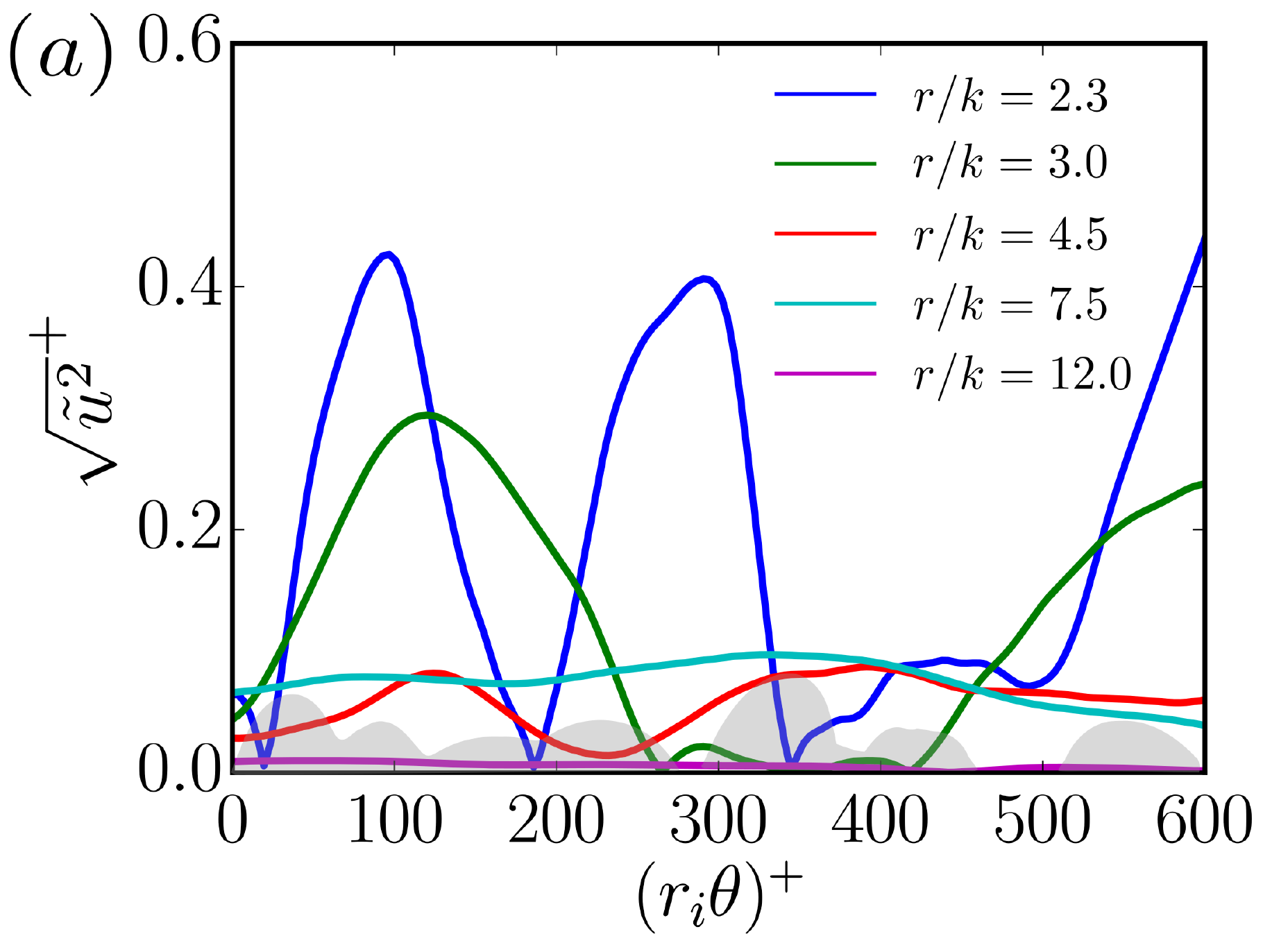}
\end{subfigure}%
\begin{subfigure}{.5\textwidth}
  \centering
  \includegraphics[width=1.0\linewidth]{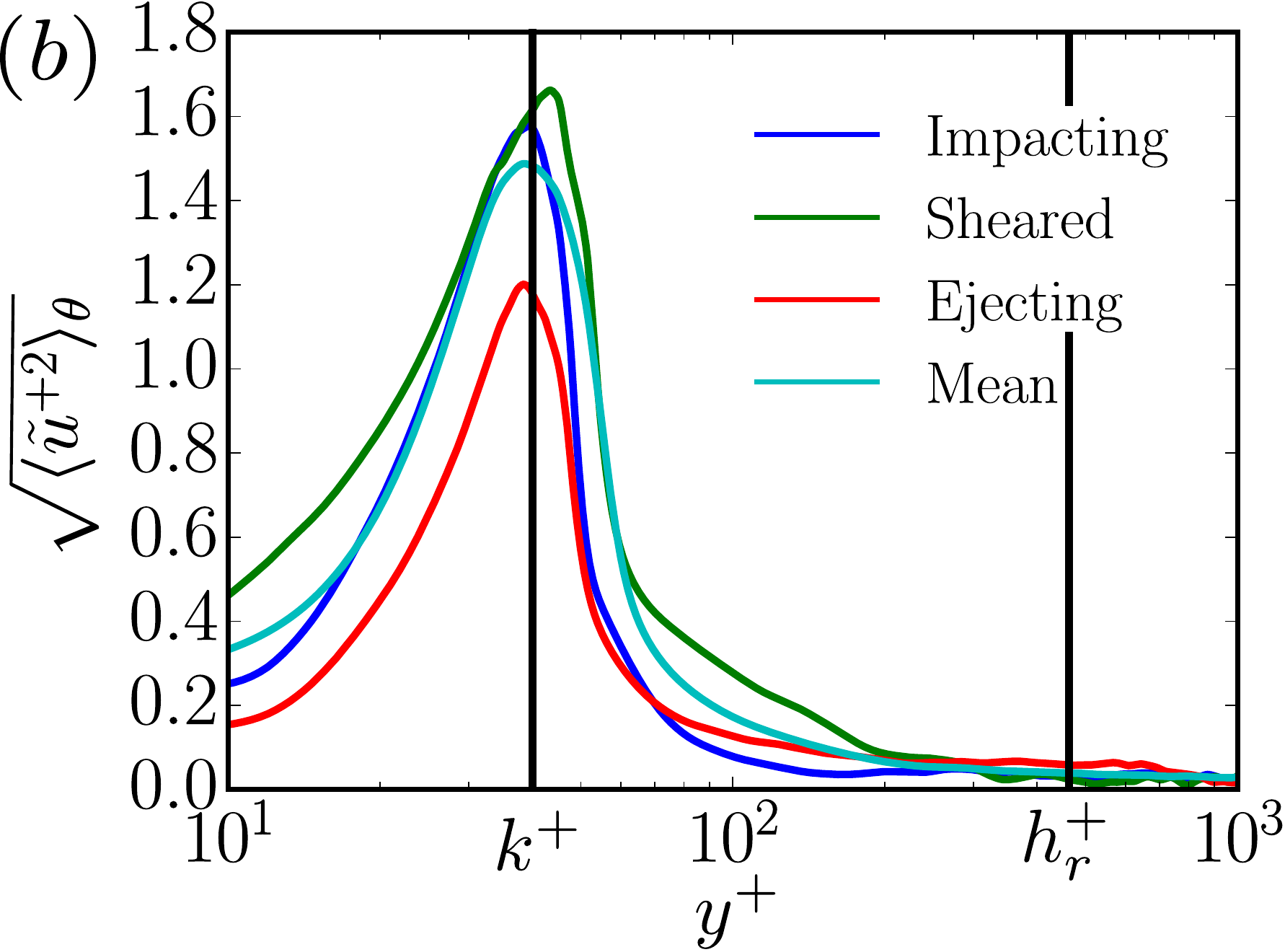}
\end{subfigure}
\caption{$(a)$ Profiles of the root mean square of the form-induced azimuthal velocity fluctuations $\tilde{u}^+ = (\bar{u} - \langle \bar{u}\rangle_\theta)/u_\tau$ at incremental heights above the roughness for identical location and conditions as in figure \ref{fig:sublayeru}. The radial coordinate $r$ is made dimensionless by the roughness height $k$. The horizontal axis represents the azimuthal coordinate in viscous units. $(b)$ Root means square of the azimuthally averaged form-induced velocity fluctuations. The profiles are obtained at three heights, namely where the Taylor roll impacts, ejects and in the center (sheared region). The cyan line gives the mean over the entire height of the cylinder. The solid black lines indicate the roughness height $k^+$ and the height of the roughness sublayer $h_r^+$.}
\label{fig:drms}
\end{figure}

A better impression of the local fluid flow disturbances induced by the roughness elements, is obtained from the time-averaged azimuthal velocity $\overline{u_\theta}$, a contour of which is shown in figure \ref{fig:sublayeru}. We zoom in on only a few roughness elements and overlay the contour with isolines of $ u_\theta^+$. It can quickly be seen that local disturbances are limited to a region of only a few times the roughness height, above which the isolines relax to approximate horizontal lines. Small recirculation zones (closed isolines) are detected in open regions behind high roughness elements. To quantify the degree of roughness-induced velocity disturbances, we apply a triple decomposition to the instanteneous azimuthal velocity $u_\theta(r,\theta,z,t)$ \citep{pokrajac07}:  
\begin{equation}
u_\theta(r,\theta,z,t)= \langle \overline{u_\theta}\rangle_\theta(r,z) + \tilde{u}_\theta(r,\theta,z) + u'_\theta(r,\theta,z,t)
\end{equation}
where $u'_\theta(r,\theta,z,t) = u_\theta(r,\theta,z,t) - \overline{u_\theta}(r,\theta,z)$ is the temporal fluctuation and $\tilde{u}_\theta(r,\theta,z)=\overline{u_\theta} (r,\theta,z) - \langle \overline{u_\theta}\rangle_\theta(r,z)$ is the component that is strongly related to the roughness induced disturbances and therefore termed the form-induced (or dispersive) velocity fluctuation (for brevity now written as $\tilde{u}$). Note that by applying the triple decomposition to the full NS equations, one will recover the related form-induced stress tensor $\langle \overline{ \tilde{u}_i\tilde{u}_j}\rangle_\theta$. However, here we only discuss $\tilde{u}$. The root mean square of the form-induced fluctuations $\sqrt{\tilde{u}^{2}}^+$ at various heights above the roughness is given in figure \ref{fig:drms}($a$). The horizontal axis corresponds to the roughness elements shown in figure \ref{fig:sublayeru}. Already for $r/k = 4.5$ (with $k$ being the roughness height) we find it hard to detect spatial  fluctuations along $\theta$. For $r/k = 12.0$, the line is barely distinguishable from the horizontal axis. If we average the lines over the azimuthal direction, we obtain the behavior of the dispersive fluctations as a function of the wall normal distance; see figure \ref{fig:drms}($b$). We plot the lines for three respective axial locations, that is the impacting, sheared and ejecting regions, and find little variance in the wall normal extent of the form-induced fluctuations, for the varying locations. The vertical black line represent the maximum roughness height, below wich we average only over fluid regions.
It is convenient to introduce a  height at which dispersive fluctuations vanish, and this height is commonly known as the `roughness sublayer height' $h_r$. In this research, we define $y^+=h_r^+$ where $\sqrt{\langle\tilde{u}^{+2} \rangle_{\theta,z}} = 0.01\langle \bar{u}^+\rangle_\theta $ and find $h_r=3.70k$ ($h_r=2.78k_s$). This value agrees well with a roughness sublayer height of $2 \lesssim k_s \lesssim 5$, as typically found in other canonical flows \citep{pokrajac07}.     
\subsection{Radial profiles}\label{profiles}
\begin{figure}
\centering
\begin{subfigure}{.5\textwidth}
  \centering
  \includegraphics[width=1.0\linewidth]{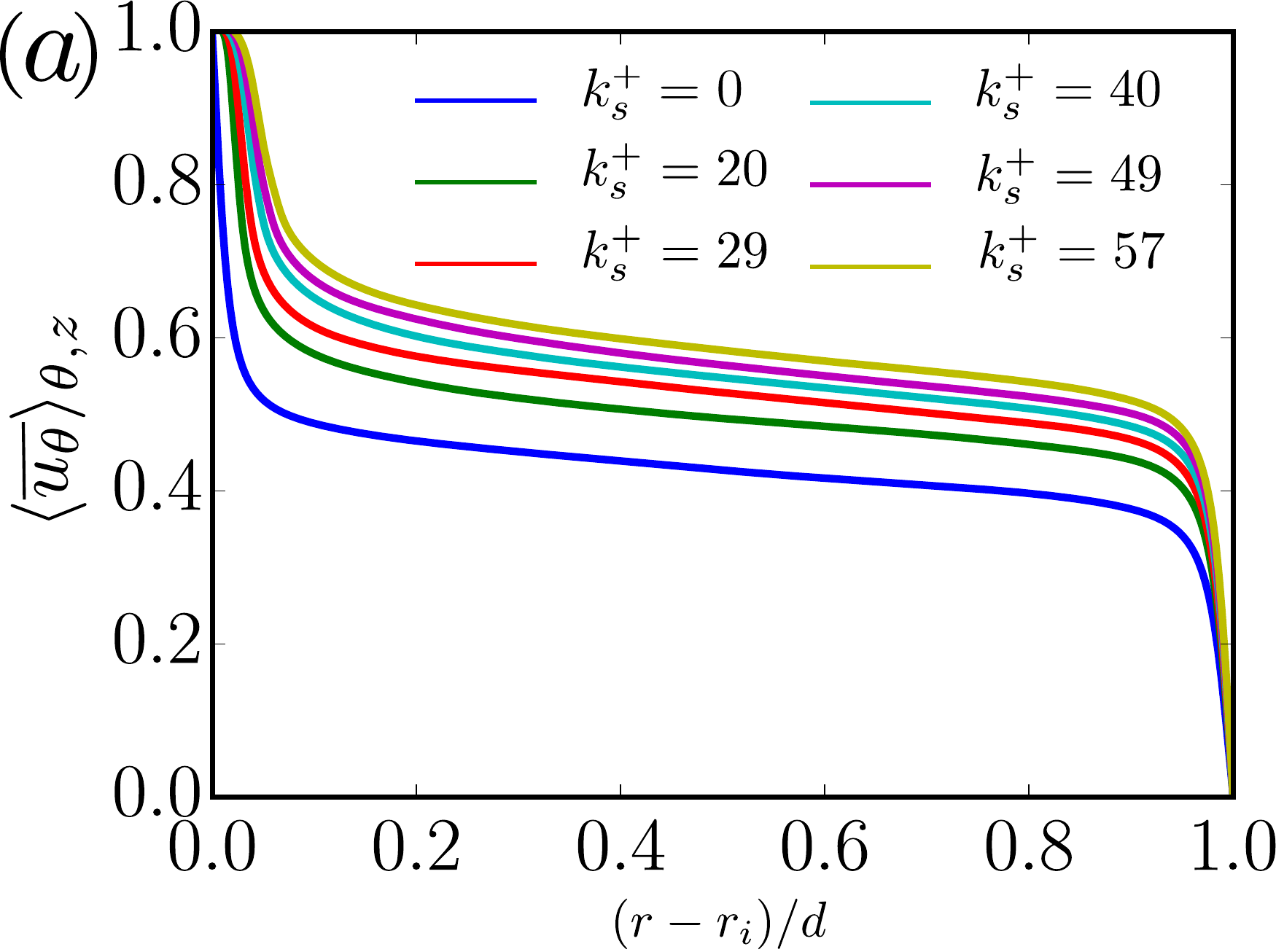}
\end{subfigure}%
\begin{subfigure}{.5\textwidth}
  \centering
  \includegraphics[width=1.0\linewidth]{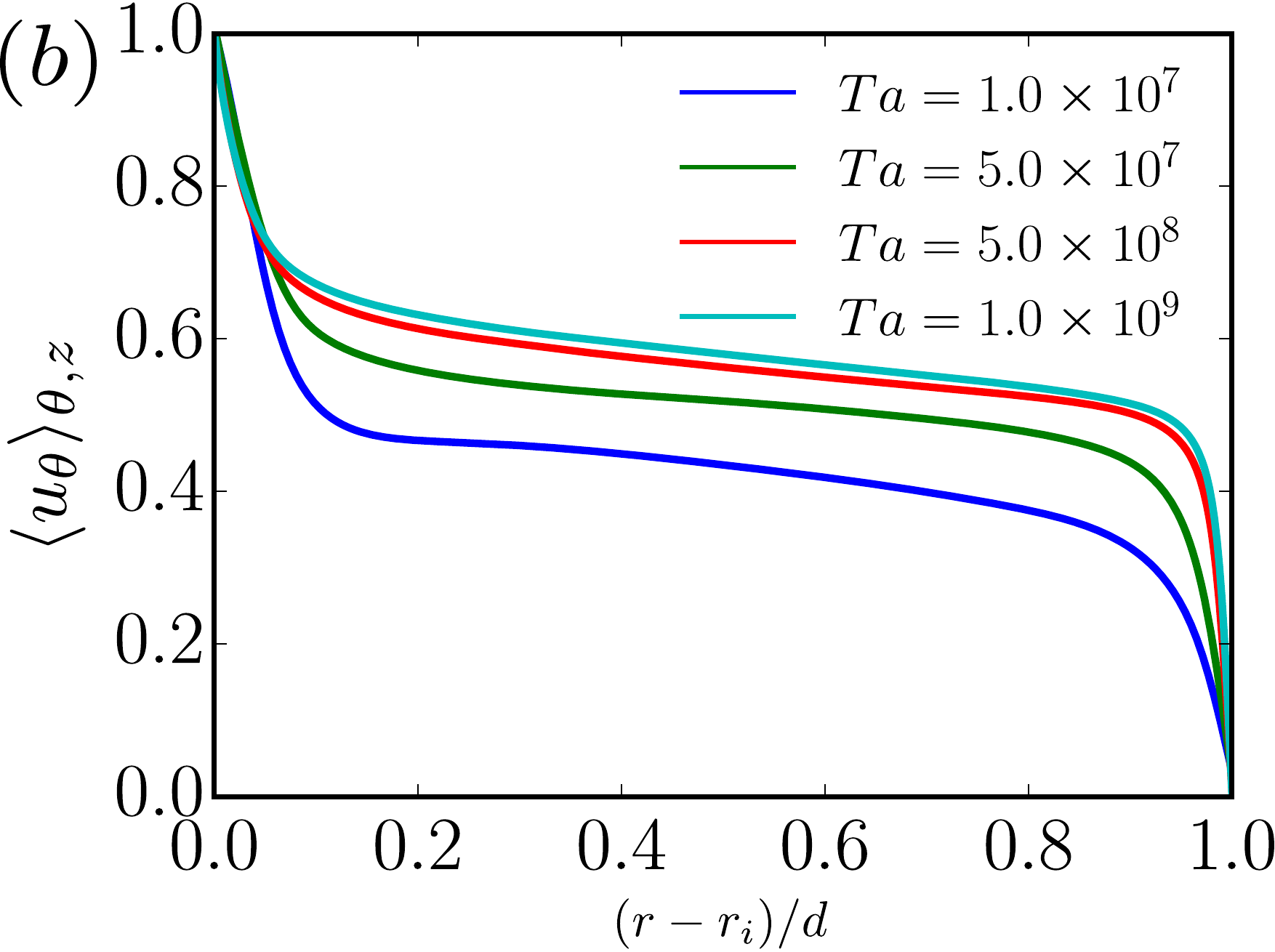}
\end{subfigure}
\caption{Profiles of the mean streamwise velocity $\langle \overline{u_{\theta}} \rangle = \langle \overline{u_{\theta}} \rangle_{\theta,z}$ versus the radial coordinate $(r-r_i)/d$, where $r_i$ the radius of the inner cylinder and $d$ is the gap width. $(a)$ Constant Taylor number $Ta=5.0\times 10^8$ and increasing roughness heights. The profiles exhibit clear `slaving', i.e. the bulk velocity moves towards the velocity of the roughened cylinder. $(b)$ Constant roughness height $k/d=0.060 \pm 0.002$ for increasing Taylor number. Increased slaving of the bulk velocity is observed for increasing viscous scaled roughness height $k_s^+$, respectively; (blue) $k_s^+ = 9$, (green) $k_s^+=24$, (red) $k_s^+= 47$ and (cyan) $k_s^+=65$.}
\label{fig:ufinal}
\end{figure}

\begin{figure}
\centering
\begin{subfigure}{.5\textwidth}
  \centering
  \includegraphics[width=1.0\linewidth]{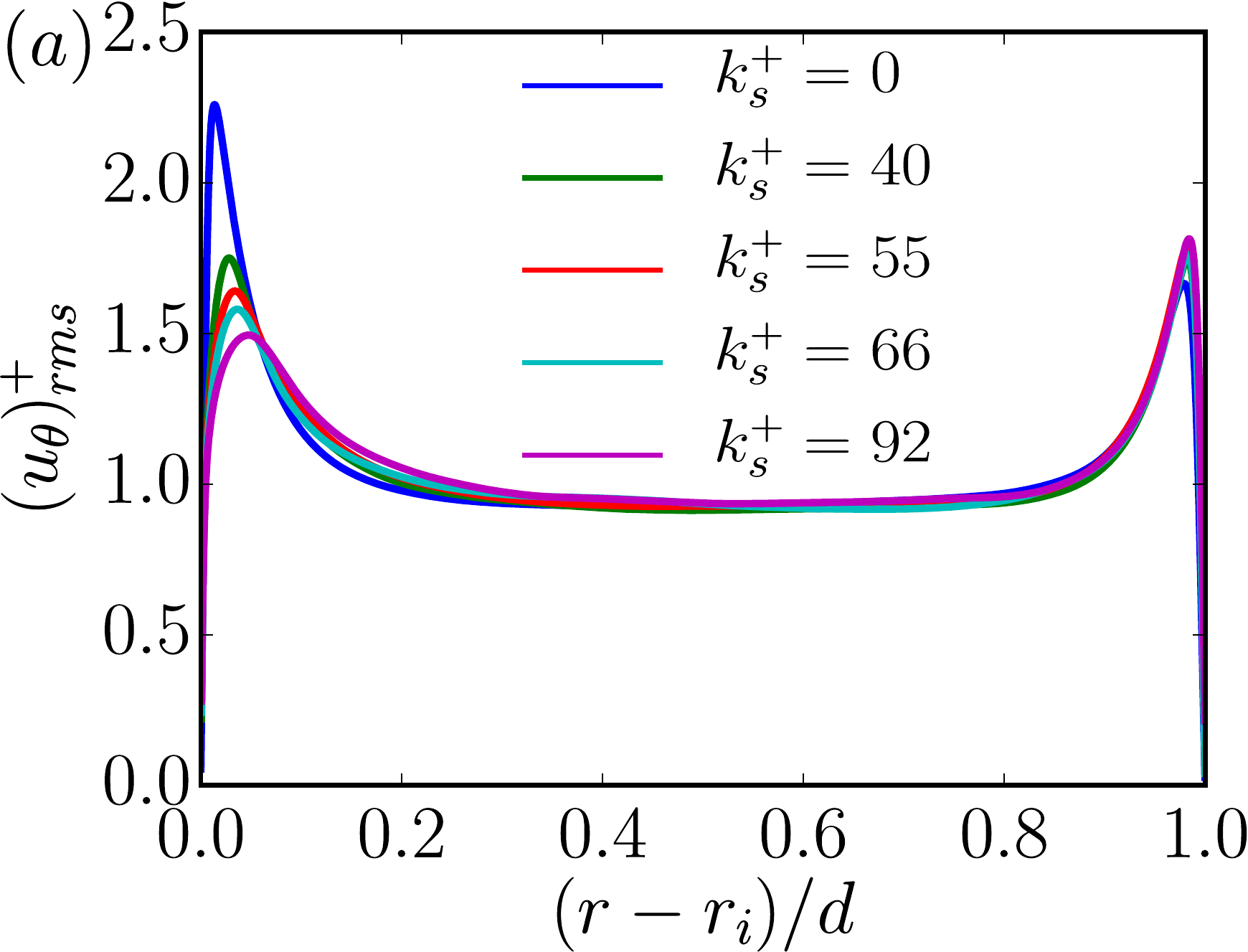}
\end{subfigure}%
\begin{subfigure}{.5\textwidth}
  \centering
  \includegraphics[width=1.0\linewidth]{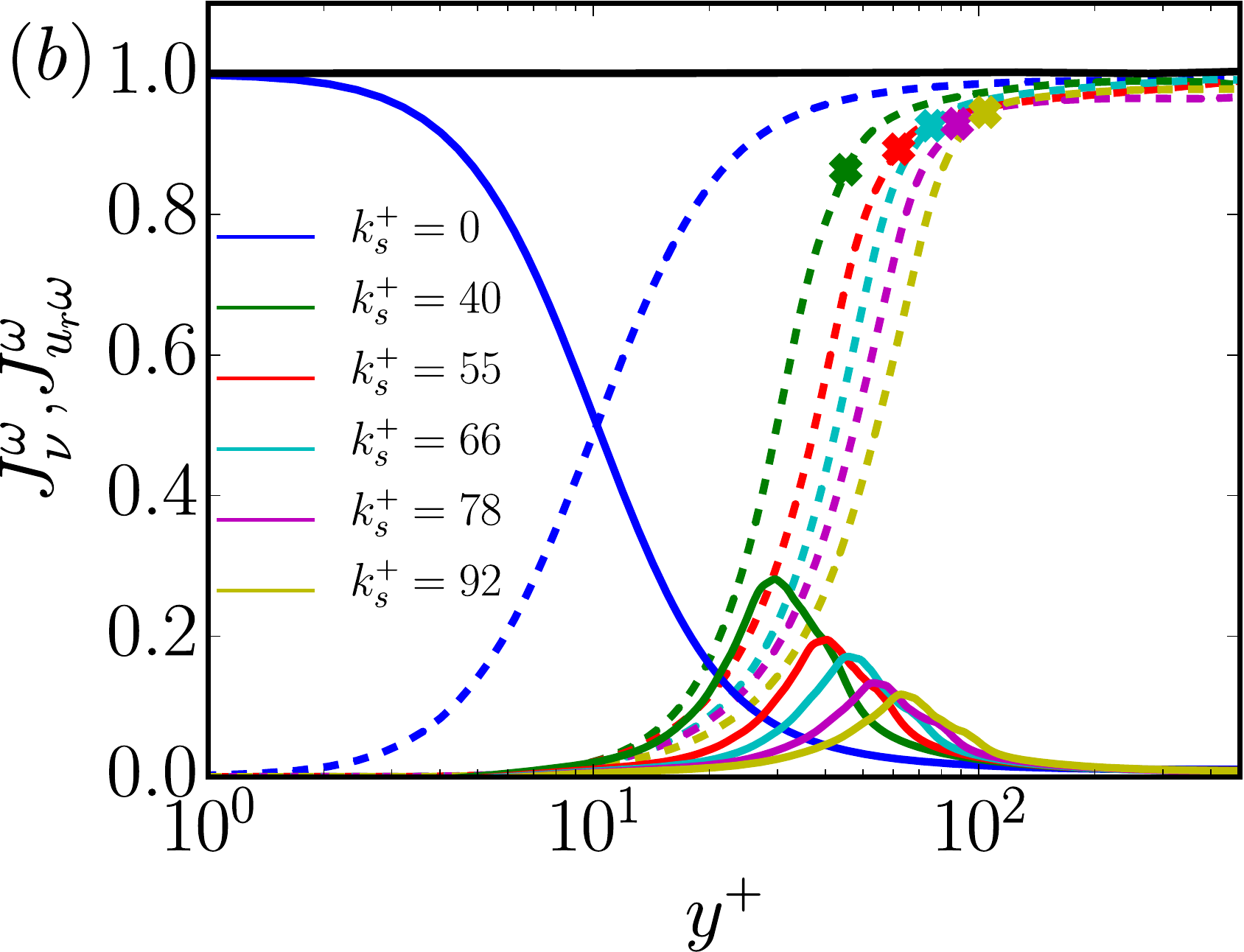}
\end{subfigure}
\caption{$(a)$ Root mean square of the mean streamwise velocity $(u_\theta)_{rms}^+ = \langle\langle \overline{u^2_\theta}\rangle_{\theta,z}-\langle \overline{u_\theta}\rangle_{\theta,z}^2\rangle_{\theta,z}^{1/2}/u_\tau$ versus the radius $(r-r_i)/d$ for $Ta=1.0\times 10^9$. We observe an overall decrease of the viscous scaled turbulence intensity for increasing roughness height $k_{s}^+$. $(b)$ The viscous $J^{\omega}_{\nu}(y^+)\equiv -\nu r^3\partial_r\langle \omega \rangle_{z,\theta,t}$ \textit{Solid line} and Reynolds stress $J^{\omega}_{u_r\omega}(y^+)\equiv -\frac{r^3}{\nu}\partial_r \langle \omega \rangle_{z,\theta,t}$ (\textit{dashed line}) terms of the conserved angular velocity flux $J^\omega(r^+)$ versus the wall normal distance $y^+$ for $Ta=1.0\times 10^9$. The sum of the individual terms represents the total conserved angular velocity flux radially outwards. The black horizontal line at $J^\omega = 1.0$ is the sum of the two blue lines for the smooth wall case. It can be seen that $J^\omega$ is conserved above the maximum roughness height (indicated by cross markers).}
\label{fig:urms}
\end{figure}
We plot the time and azimuthally averaged azimuthal velocity for $Ta=5.0\times 10^8$ in figure \ref{fig:ufinal}($a$). The roughness covers the inner cylinder, i.e. $(r-r_i)/d \le 0.07$. Below the roughness crest, the velocity is averaged over the entire domain, including the solid rough regions. For increasing roughness height, see the legend for viscous roughness heights, the azimuthal velocity in the bulk increases. Figure \ref{fig:ufinal}($b$) then presents the corresponding azimuthal velocity profiles for constant $k/d = 0.060 \pm 0.002$ and varying Taylor number. 

Figure \ref{fig:urms}($a$) shows the double-averaged radial profiles $(u_{\theta})_{rms}^+ = \langle\langle \overline{u^2_\theta}\rangle_{\theta,z}-\langle \overline{u_\theta}\rangle_{\theta,z}^2\rangle_{\theta,z}^{1/2}/u_{\tau}$ for constant Taylor number $Ta=1.0\times 10^9$. For the smooth wall case, the peak of the rms of the azimuthal velocity is located at $u_{\theta,rms}|_{max}\approx 10y^+$. This agrees with the smooth case in \citep{ostilla16} for TC flow, but is closer to the solid boundary than that is found in channel flow; $u_{\theta,rms}|_{max}\approx 12y^+$. We observe a slight decrease in the viscous scaled turbulence intensity $(u_{\theta})_{rms}^+$ for increasing roughness heights, close to the inner cylinder. Further outside the profiles, the profiles collapse. This is once more a sign that the effects of roughness are only confined to the fuid flow close to the wall.

It has already been mentioned that the angular velocity flux $J^\omega(y^+)$ is conserved in the radial direction, see section \ref{global}. However, the individual components -- i.e. the viscous $J_{\nu}^\omega(y^+)$ and Reynolds stress $J_{u_r\omega}^\omega(y^+)$ terms -- are functions of the wall normal (radial) coordinate. The profiles of these terms are shown in figure \ref{fig:urms}(b). We normalize all terms with $max(J^\omega(y^+))$. The viscous stress terms are presented as solid lines and the Reynolds stress terms as dashed lines. The blue lines represent the smooth wall case. Very close to the inner cyliner ($y^+=1$), $J^\omega \approx J_{\nu}^\omega$; this is expected, since there the gradient of the mean streamwise velocity is maximum and the wall normal component of the velocity vector goes to zero. On the far right of figure \ref{fig:urms}(b), the gradient of the mean velocity approaches zero in the bulk, and the correlation function between the wall normal and angular velocity goes to a maximum; as such $J^\omega \approx J_{u_r\omega}^\omega$. The black line represents the sum of the viscous and Reynold stress terms, for the smooth (blue) case, and is independent of $y^+$. The situation for the rough cylinder cases is more complex. For increasing roughness height, the viscous stress reaches a maximum below the maximum roughness height. This can be explained by the recirculation zones behind the roughness elements. Then, above the roughness, the viscous stress goes to zero in the bulk. For the Reynolds stress terms the increase is monotonic, and similar to the smooth wall case. However, we observe a steeper increase for the rough cases. Note that under the maximum roughness height, the viscous and Reynolds stress terms do not add up to unity, since the rough surface acts as a radial dependent momentum source term to the NS equations there. 

\section{Summary \& conclusions}\label{conclusions}
We have performed Direct Numerical Simulations of turbulent Taylor--Couette flow with inner cylinder rotation and inner cylinder sand grain roughness. The Taylor number ranges from $Ta=1.0\times 10^7$ ($Re_\tau = 82$) up to $Ta=1.0\times 10^9$ ($Re_\tau = 635$), covering thereby both the classical and the ultimate regimes of turbulent TC flow. In particular, we studied the effects of the roughness height on the fluid flow in the transitionally rough and fully rough regimes, with the equivalent sand grain roughness height ranging from $k_s^+= 5$ to $k_s^+=92$. We modelled the sand grains as randomly rotated and translated ellipsoids of constant size and shape (\textit{monodisperse}), similar to the model proposed by \cite{Scotti06}. The surface was implemented into a second order finite difference code by means of the Immersed Boundary Method.

We confirm an increase in the dimensionless torque, expressed as the Nusselt number, for increasing roughness height. This is attributed to the enhanced boundary layer detachment, resulting in plume ejection regions, which we observed in snapshots of the azimuthal velocity field. The plumes contain a strong radial velocity component, and as such contribute strongly to the Reynolds stress term of the angular velocity flux. This mechanism is analogous to what was found for the Nu increase for grooved TC turbulence by \cite{zhu16}. 

To quantify the degree of roughness induced disturbances to the velocity field, we measured the dispersive fluctuations of the azimuthal velocity component, $\tilde{u}_\theta = \langle u_\theta \rangle_t - \langle u_\theta \rangle_{t,\theta}$. This dispersive term was obtained by means of a triple decomposition to the azimuthal velocity. We defined the height of the roughness sublayer $h_r$ there where the dispersive fluctuations become very small, such that $\sqrt{\langle\tilde{u}^{+2} \rangle_{\theta,z}} = 0.01\langle \bar{u}^+\rangle_\theta $ and found the height of the roughness sublayer to be $h_r = 2.78k_s$. This height of the roughness sublayer compares well with values found for other canonical systems \citep{pokrajac07}.

The hallmark of turbulent flows over rough walls is the shift of the logarithmic streamwise velocity profile $\Delta u_\theta^+$. The shift is a function of any parameters describing the roughness topology. Here we focused in particular on the effect of the sand grain size $k$ and the roughness function becomes $\Delta u_\theta^+ = f(k)$. It was shown in \cite{huisman13} that the constants of the logarithmic law are not constant in the Taylor number range of our simulations. Hence we proposed the generalization; $u_\theta^+ = f_1(Re_i)\log(y^+)+f_2(Re_i) - \Delta u_\theta^+$.

We performed simulations at four Ta and various roughness heights and ensured that the $k_s^+$ range for the various Ta numbers overlaps. First, we concluded that the velocity shift is independent of Ta, despite the Ta dependence of the constant in the logarithmic layer. As such, all simulations collapse onto a single curve. Second, we saw a strong overlap between the roughness function calculated from our DNS in TC flow and the seminal work by \cite{Nikuradse33} on \textit{monodisperse} sand grains in pipe flow, in the transitionally rough regime. We found $k_s^+=1.33k$.  Only for very low $k_s^+$ values, close to the hydrodynamically smooth regime, we found that the simulations slightly differ from the Nikuradse curve. 

It is remarkable that the Hama roughness function appears to be universal for similar surfaces in such different systems. Note in particular that we have a streamwise curvature and strong secondary motions (Taylor rolls), which were absent in the pipe flow experiments of Nikuradse. As such, our findings point towards a universal behavior of the roughness function for very different fluid flow systems. However, many more comparison studies, of identical rough surfaces and varying fluid flow systems, are needed to confirm this notion.

\section*{Acknowledgements}
We wish to express our gratitude to Jelle Will for his help with various illustrations. This project is funded by the 
Priority Programme SPP 1881 Turbulent Superstructures of the Deutsche Forschungsgemeinschaft. We also acknowledge PRACE for awarding us access to Marconi, based in Italy at CINECA under PRACE project number 2016143351. We also acknowledge PRACE for awarding us access to MareNostrum based in Italy at the Barcelona Supercomputing Center (BSC) under PRACE project number 2017174146. This work was partly carried out on the national e-infrastructure of SURFsara, a subsidiary of SURF cooperation, the collaborative ICT organization for Dutch education and research.

\appendix
\section{}\label{appA}
\begin{table}
\centering
\begin{tabular}{l|llllllllllllllll}
Case & $\bar{k}/k_{max}$ & $\bar{k}/k$ & $k_{std}/k$ & Sk & Ku & $k_{rms}/k$ & $k_p/k$ & ES & $\alpha_{rms}$ & $S_f/r_i^2$ & $S_s/r_i^2$ & $\Lambda$ \\
\multicolumn{8}{c}{} \\
A1 & $0.38$     &$0.57$ & $0.35$ & $-0.06$ & $-0.77$ & $0.67$ & $1.11$ & $0.44$ & $0.17$ & $0.15$ & $0.21$ & $9$ \\
A2 & $0.38$     & $0.57$ & $0.35$ & $-0.08$ & $-0.78$ & $0.67$ & $1.11$ & $0.44$ & $0.17$ & $0.16$ & $0.21$ & $9$ \\
A3 & $0.37$     & $0.56$ & $0.34$ & $-0.09$ & $-0.76$ & $0.66$ & $1.42$ & $0.46$ & $0.11$ & $0.17$ & $0.24$ & $9$  \\
A4 & $0.37$     & $0.56$ & $0.36$ & $-0.11$ & $-0.87$ & $0.67$ & $1.05$ & $0.48$ & $0.12$ & $0.18$ & $0.28$ & $10$  \\
\multicolumn{8}{c}{} \\
B1 & $0.38$ & $0.57$ & $0.35$ & $-0.06$ & $-0.77$ & $0.67$ & $1.11$ & $0.44$ & $0.17$ & $0.15$ & $0.21$ & $9$   \\
B2 & $0.38$ & $0.57$ & $0.35$ & $-0.08$ & $-0.78$ & $0.67$ & $1.12$ & $0.44$ & $0.17$ & $0.16$ & $0.21$ & $9$   \\
B3 & $0.37$ & $0.56$ & $0.34$ & $-0.09$ & $-0.76$ & $0.66$ & $1.04$ & $0.46$ & $0.11$ & $0.17$ & $0.24$ & $9$   \\
BY & $0.38$ & $0.57$ & $0.35$ & $-0.07$ & $-0.75$ & $0.67$ & $1.07$ & $0.47$ & $0.15$ & $0.17$ & $0.25$ & $9$   \\
B4 & $0.37$ & $0.56$ & $0.36$ & $-0.11$ & $-0.87$ & $0.67$ & $1.05$ & $0.48$ & $0.12$ & $0.18$ & $0.28$ & $10$   \\
B5 & $0.37$ & $0.56$ & $0.35$ & $-0.13$ & $-0.86$ & $0.66$ & $1.02$ & $0.44$ & $0.11$ & $0.17$ & $0.26$ & $9$   \\
\multicolumn{8}{c}{} \\
C1 & $0.38$ &$0.57$ & $0.35$ & $-0.08$ & $-0.81$ & $0.67$ & $1.07$ & $0.47$ & $0.13$ & $0.17$ & $0.24$ & $9$  \\ 
C2 & $0.37$ &$0.56$ & $0.35$ & $-0.09$ & $-0.85$ & $0.66$ & $1.03$ & $0.47$ & $0.11$ & $0.16$ & $0.24$ & $10$ \\
C3 & $0.37$ &$0.56$ & $0.36$ & $-0.11$ & $-0.85$ & $0.67$ & $1.04$ & $0.48$ & $0.07$ & $0.18$ & $0.29$ & $10$ \\
C4 & $0.37$ &$0.56$ & $0.35$ & $-0.10$ & $-0.80$ & $0.66$ & $1.04$ & $0.47$ & $0.12$ & $0.17$ & $0.25$ & $10$ \\
C5 & $0.37$ &$0.56$ & $0.35$ & $-0.10$ & $-0.82$ & $0.67$ & $1.04$ & $0.48$ & $0.12$ & $0.17$ & $0.26$ & $10$  \\
\multicolumn{8}{c}{} \\
D1 & $0.37$ & $0.56$ & $0.35$ & $-0.09$ & $-0.85$ & $0.66$ & $1.03$ & $0.47$ & $0.11$ & $0.16$ & $0.24$ & $10$ \\
D2 & $0.37$ & $0.56$ & $0.36$ & $-0.11$ & $-0.85$ & $0.67$ & $1.03$ & $0.48$ & $0.06$ & $0.19$ & $0.30$ & $10$ \\
D3 & $0.37$ & $0.56$ & $0.35$ & $-0.11$ & $-0.82$ & $0.66$ & $1.01$ & $0.48$ & $0.08$ & $0.17$ & $0.27$ & $11$ \\
D4 & $0.37$ & $0.56$ & $0.36$ & $-0.10$ & $-0.83$ & $0.67$ & $1.02$ & $0.49$ & $0.08$ & $0.18$ & $0.29$ & $11$ \\
D5 & $0.37$ & $0.56$ & $0.35$ & $-0.12$ & $-0.85$ & $0.67$ & $1.01$ & $0.50$ & $0.08$ & $0.19$ & $0.32$ & $10$ \\
\end{tabular}
\caption{Parameters describing the surface geometry. From left to right; Case number in accordance with table \ref{table:sim_param}. The roughness height is measured with respect to the inner cylinder location $r=r_0$. $\bar{k}/k_{max}$ is the mean roughness height respective to the maximum roughness height. $\bar{k}$ is the first moment of the roughness height distribution; $\frac{1}{S}\int_{S} k(z,\theta) dS$. $k_{std}$ is the standard deviation of the surface height distribution $(\frac{1}{S}\int_{S} (k(z,\theta)-\bar{k})^2dS)^{1/2}$, Sk is the skewness $\frac{1}{k_{std}^3S}\int_{S} (k(z,\theta)-\bar{k})^3dS$ and Ku the kurtosis, $\frac{1}{k_{std}^4S}\int_{S} (k(z,\theta)-\bar{k})^4dS$. $k_{rms}$ is the root mean square of the mean height $(\frac{1}{S}\int_{S} (k(z,\theta))^2dS)^{1/2}$. $k_p$ is the mean peak height. The peaks are obtained by a peak finding algorithm (`\textit{find\_peak\_local}' in \cite{python_image}) with zero threshold and a minimum spacing of the peaks of 4 grid nodes. ES is the effective slope, $ES=\frac{1}{L_\theta L_z}\int_{L_\theta} \int_{L_z} |\frac{\partial k(z,\theta)}{\partial (r_i\theta)}|dzd(r_i\theta)$ as introduced by \cite{napoli08}. It can be shown that the ES parameter is twice the often used solidity parameter $\lambda$ \citep{jimenez04}. $\alpha$ is the surface gradient in the streamwise direction, $\frac{\partial k(z,\theta)}{\partial \theta}$,  and $\alpha_{rms}$ is the root mean square of this distribution. $S_f$ is the total frontal projected area of the roughness (in the streamwise direction). $S_s$ is the total windward wetted area of the roughness.  $\Lambda$ is the density parameter which relates the latter two parameters by $\Lambda = (\frac{S}{S_f})(\frac{S_f}{S_s})^{-1.6}$ where $S$ is the total area of the surface without roughness \citep{sigal90}. The fractal dimension of surface $B2$ is calculated from the slope of the power density spectrum versus the wavenumber, $C \propto q^{-4.65}$, for details we refer to \citep{persson05}. Here we find a Hurst exponent $H$ of $1.32$ and a fractal dimension $D$ of $1.68$. Such we conclude that the surface can at no scale be considered fractal.}
\label{table:surface}
\end{table}
\clearpage

\bibliographystyle{jfm}
\bibliography{jfm-instructions}
\end{document}